\documentclass[letterpaper]{article} 
\usepackage{aaai2026}
\nocopyright
\usepackage{times}  
\usepackage{helvet}  
\usepackage{courier}  
\usepackage[hyphens]{url}  
\usepackage{graphicx} 
\usepackage{natbib}  
\usepackage{caption} 
\usepackage{algorithm}
\usepackage{algorithmic}
\usepackage{booktabs}
\usepackage[table]{xcolor}
\usepackage{makecell}
\usepackage{multirow}
\usepackage{amsmath}
\usepackage{amssymb}
\usepackage{newfloat}
\usepackage{listings}
\DeclareCaptionStyle{ruled}{labelfont=normalfont,labelsep=colon,strut=off} 
\floatstyle{ruled}
\newfloat{listing}{tb}{lst}{}
\floatname{listing}{Listing}
\title{Beyond Interactions: Node-Level Graph Generation for Knowledge-Free Augmentation in Recommender Systems}
\author{
    Zhaoyan Wang\thanks{Corresponding author.},
    Hyunjun Ahn,
    In-Young Ko
}
\affiliations{
    School of Computing, KAIST, Daejeon, Republic of Korea\\
    \{zhaoyan123, a.hyunjun, iko\}@kaist.ac.kr
}

\usepackage{bibentry}

\begin{document}

\maketitle

\begin{abstract}
Recent advances in recommender systems rely on external resources such as knowledge graphs or large language models to enhance recommendations, which limit applicability in real-world settings due to data dependency and computational overhead. Although knowledge-free models are able to bolster recommendations by direct edge operations as well, the absence of augmentation primitives drives them to fall short in bridging semantic and structural gaps as high-quality paradigm substitutes. Unlike existing diffusion-based works that remodel user-item interactions, this work proposes NodeDiffRec, a pioneering knowledge-free augmentation framework that enables fine-grained node-level graph generation for recommendations and expands the scope of restricted augmentation primitives via diffusion. By synthesizing pseudo-items and corresponding interactions that align with the underlying distribution for injection, and further refining user preferences through a denoising preference modeling process, NodeDiffRec dramatically enhances both semantic diversity and structural connectivity without external knowledge. Extensive experiments across diverse datasets and recommendation algorithms demonstrate the superiority of NodeDiffRec, achieving State-of-the-Art (SOTA) performance, with maximum average performance improvement 98.6\% in Recall@5 and 84.0\% in NDCG@5 over selected baselines.
\end{abstract}


\section{Introduction}
\begin{figure}[t]
\centering
\includegraphics[width=\linewidth]{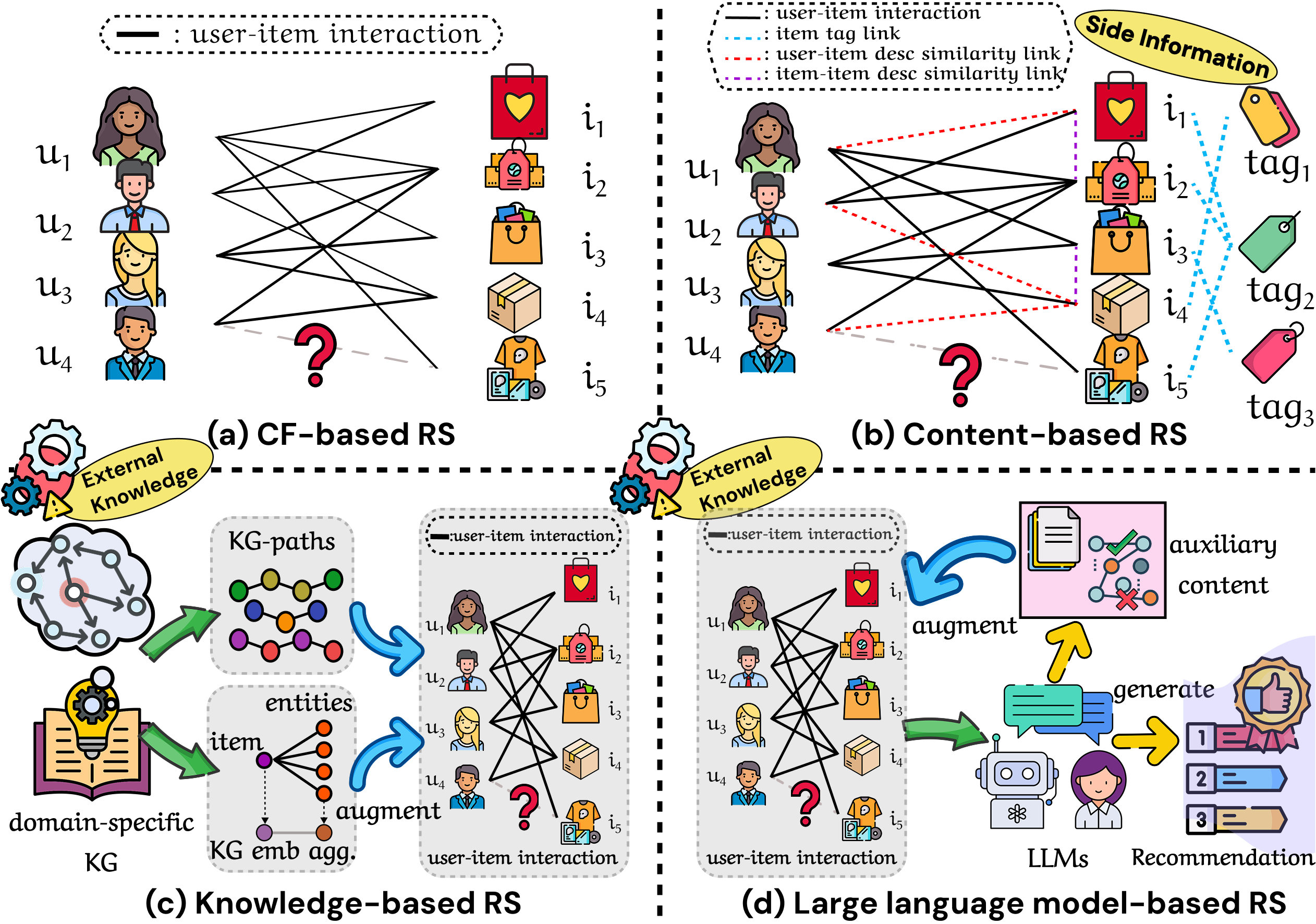}
\caption{Current recommendation paradigms: (a) CF-based: analyzes patterns across similar users or items. (b) Content-based: recommends items by content features. (c) Knowledge-based: recommends items based on external domain knowledge. (d) LLM-based: employs LLMs to understand users' preferences through natural language reasoning.}
\label{Overview Figure}
\end{figure}

Recommender systems~\cite{fan2019graph, ying2018graph, van2013deep, covington2016deep, schafer2001commerce}, designed as intelligent filters for personalized content, have become a central focus in both industry and academic research~\cite{gao2023survey}. With recent developments, current personalized recommender systems can be classified into four core categories: \textbf{Collaborative Filtering (CF)-based}, \textbf{Content-based}, \textbf{Knowledge-based}, and \textbf{Large Language Model (LLM)-based} recommender systems~\cite{li2024recent}, illustrated in Figure \ref{Overview Figure}. Besides, \textbf{hybrid} approaches are also widely adopted to leverage the strengths of each.

In practice, to address the longstanding sparsity and cold-start challenges, LLM- and knowledge-enhanced recommender systems have demonstrated superiority and become two prominent augmentation paradigms~\cite{lin2025can,chicaiza2021comprehensive,zhang2024review}.

Knowledge-enhanced paradigms leverage structured domain knowledge such as Knowledge Graphs (KGs), through semantic reasoning and external information integration, where embedding-based techniques and semantic paths are often exploited~\cite{meng2025doge,jiang2024diffkg,yang2022knowledge,elahi2024knowledge,yang2024sequential}. Meanwhile, LLM-enhanced ones augment semantic understanding and user-item interaction by enabling natural-language explanations, modifying user or item embeddings, and contributing to semantic retrieval or re-ranking~\cite{meng2025doge,xi2024towards,sun2025llm4rsr,qiu2021u,liu2025cora}.

While knowledge- and LLM-enhanced paradigms provide viable augmentation frameworks, there are several inherent limitations introduced, which severely restrict their applicability in practical, real-world environments:

1) Both KG- and LLM-based augmentation heavily rely on extensive data and domain knowledge. In cases where such knowledge is unavailable or where KGs and LLMs are infeasible to be deployed due to computational and access constraints, augmentation cannot be performed.

2) For KGs, high construction and maintenance costs, along with data sparsity and noise issues, limit the reliability and scalability of KG-enhanced methods~\cite{guo2020survey}.

3) For LLMs, their intensive resource and computational demands, difficulty in controllable generation, and the risk of hallucination collectively hinder their deployment and benefits in recommender systems~\cite{lin2025can}.

Such limitations highlight the need for self-contained augmentation strategies with no dependence on external domain knowledge and computationally intensive models, generating auxiliary signals from intrinsic patterns within given datasets in a self-sufficient manner. As a result, knowledge-free generative models such as variational auto-encoder (VAEs), auto-regressive models, generative adversarial networks (GANs), and diffusion models (DMs) may serve as an ideal alternative source of augmentation data to the external~\cite{kingma2013auto,van2016pixel,sohl2015deep}.

However, knowledge-free approaches still remain fundamentally constrained, due to the critical research gap resulting from absent augmentation primitives: the incapability to incorporate new entities like how the KG-based operate, and failing to emulate LLM-enhanced methods in enriching embedding spaces semantically, while preserving logical relationships between augmented content and the original.

\begin{figure}[t]
\centering
\includegraphics[width=\linewidth,]{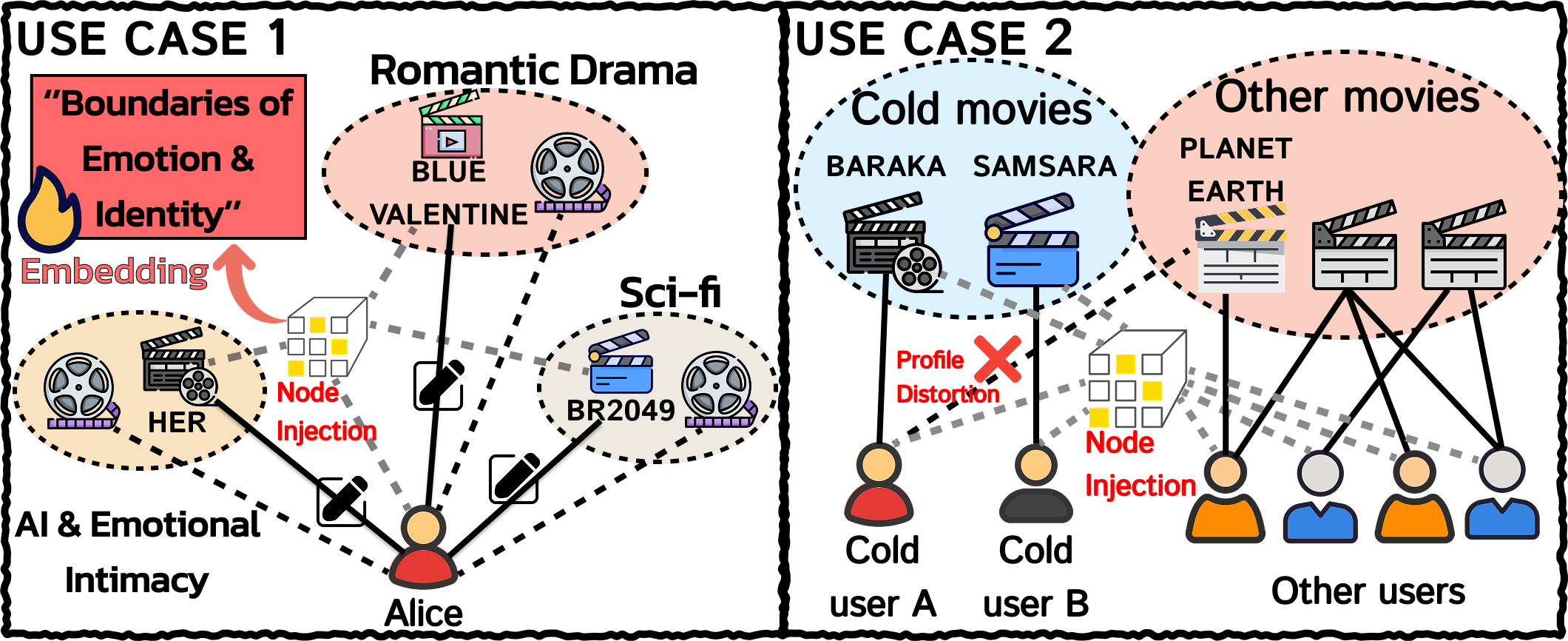}
\caption{Constraints of SOTA knowledge-free model-based augmentation: (1) Interest Expression Gaps in Embedding Space; (2) Structural Instability in Isolated Cold Entities.}
\label{use cases}
\end{figure}

Specifically, SOTA knowledge-free generators, diffusion models, enhance recommendation by user-item interaction deletion, creation, and reweighting operations~\cite{liu2023diffusion,walker2022recommendation,he2024diffusion,wang2023diffusion}, exhibiting inherent flaws along both semantic and structural dimensions. Illustrated by Case 1 in Figure \ref{use cases}, such operations solely with observed items cannot synthesize latent semantic regions in the embedding space, leading to under-representation of user interests. For instance (Use case 1), users with complex, cross-genre movie preferences may not have proper movies reflecting their holistic interests. In Case 2, cold entities remain structurally disconnected, where edge modifications are insufficient to establish informative context without profile distortion. When movie items have limited play history, they suffer from unstable embeddings due to the lack of contextual grounding. Direct edge operations obscure the true movie types that cold users prefer (Use case 2). Appendix Section~\ref{Use case Analysis} provides more detailed illustrations.

The above two cases demonstrate that current primitives of knowledge-free generators fall short in bridging semantic and structural gaps as high-quality paradigm substitutes, motivating the necessity for expressive and topology-aware augmentation primitive expansion. Node injection addresses the challenges in both use cases directly: it enables semantic coverage of latent user interests via synthetic nodes, and provides structurally meaningful bridges for cold entities.

Therefore, we suggest that developing the absent entity-injection primitive through graph generation, along with edge augmentation centered on generated entities, matures knowledge-free recommendation augmentation and offers a promising paradigm without external knowledge. However, current studies on graph generation focus on graph-level tasks, where the output is generated graphs resembling the given graphs, but not specific nodes and edges on purpose. This poses pivotal challenges for user-item optimization that requires precise node- and edge-level operations.

To address the above issues, we propose \textbf{NodeDiffRec}, the first \underline{\textbf{Node}}-level \underline{\textbf{Diff}}usion Model for \underline{\textbf{Rec}}ommendation, which generates new items with corresponding user-item interactions through an injection diffusion process, and places them properly to conduct node-level augmentation, specifically designed for recommendation. NodeDiffRec eliminates the accompanying structural noise by another preference modeling diffusion step. It substantially fills the gap of missing fine-grained augmentation primitives in previous knowledge-free generators via graph generation, boosting recommendation performance with no external knowledge. Our contributions can be summarized as follows:
\begin{itemize}
    \item We significantly enhance recommendations without any external knowledge by proposing NodeDiffRec, a novel two-stage injection-denoising diffusion framework.
    \item To the best of our knowledge, NodeDiffRec is the first work to introduce graph generation techniques adapted to the node level for nuanced generation in recommender systems, while other DMs model user-item interactions.
    \item We successfully supplement the missing entity-injection and injection-centered edge augmentation primitives for knowledge-free generators, laying the foundation for future research on self-contained augmentation paradigms.
    \item Extensive experiments on three recommendation datasets validate NodeDiffRec's superiority over existing generative models, achieving SOTA augmentation performance.
\end{itemize}

\begin{figure*}[t]
\centering
\includegraphics[width=\linewidth,height=6cm]{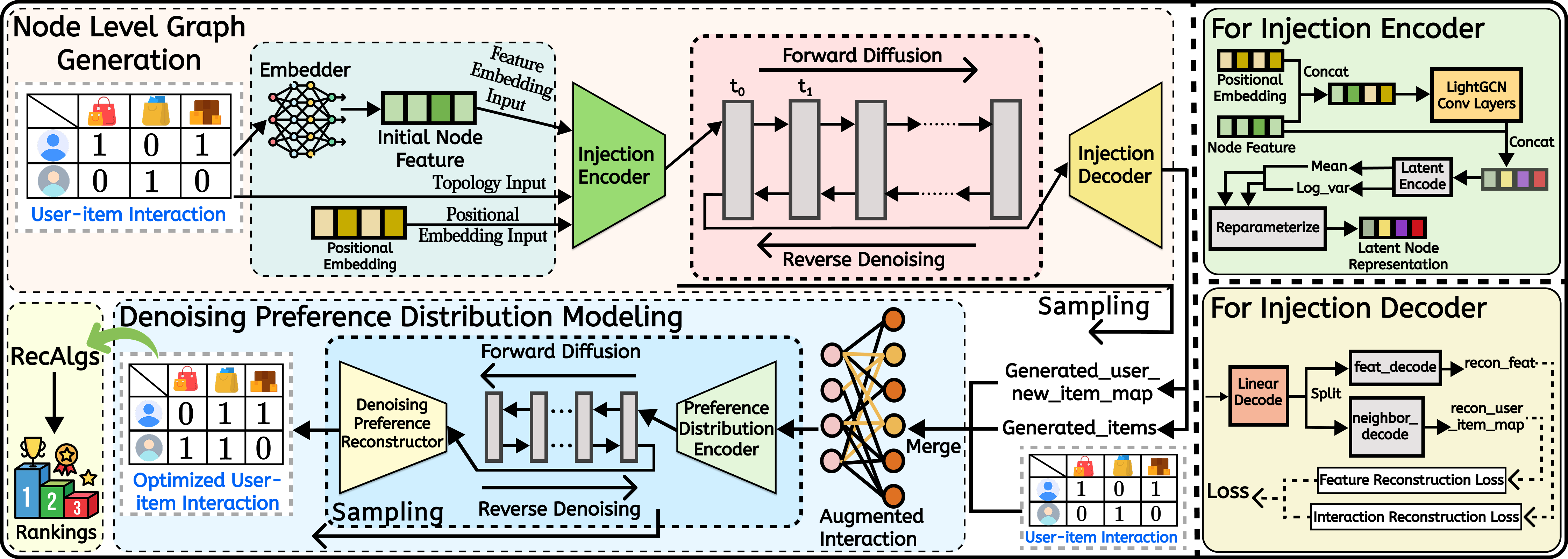}
\caption{The architecture of NodeDiffRec, consisting of two diffusion phases: a node-level graph generation process for expanded nuanced injection and interaction augmentation operations, followed by another denoising preference modeling phrase.}
\label{system overview}
\end{figure*}

\section{Related Work}
\subsection{Recommendation Algorithms}\label{sec:Recommendation Algorithm}
Recommendation algorithms have progressed from matrix factorization (MF) to deep and graph-based models to better capture complex interactions~\cite{roy2022systematic}. Among early approaches, Alternating Least Squares (ALS)~\cite{hu2008collaborative} and PureSVD~\cite{cremonesi2010performance} learn user and item embeddings by projecting sparse interactions into a low-dimensional space. Building on this, Neural Matrix Factorization (NeuMF)~\cite{he2017neural} incorporates neural networks to model non-linear preference patterns, achieving better expressiveness. Later, Neural Graph Collaborative Filtering (NGCF)~\cite{wang2019neural} extends modeling to graph structures, propagating user–item signals over neighborhoods. In self-supervised learning, methods such as Self-supervised Graph Learning (SGL)~\cite{wu2021self} and SimGCL~\cite{yu2022graph} improve representation robustness by contrastive learning.

\subsection{Graph Generation Techniques}
Graph generation techniques exhibit broad applications in molecule design and protein modeling tasks. Approaches based on VAEs, GANs, and normalizing flows have shown progress~\cite{jin2018junction,de2018molgan,luo2021graphdf}, yet they struggle with data discreteness and complex structural dependencies. Motivated by diffusion models' success in the computer vision domain, originally developed for image generation, recent studies have introduced DMs for graph generation. Representative paradigms include score matching with Langevin dynamics, denoising diffusion probabilistic models, and score-based models~\cite{liu2023generative}. Models such as EDP-GNN, DiGress, and GDSS~\cite{niu2020permutation,vignac2022digress,jo2022score} achieved promising outcomes in molecular and structural graph generation.

However, current works predominantly focus on graph-level generation, aiming to learn the distribution and synthesize entire new graphs, overlooking the substantial challenges to more granular and controllable generation.

\subsection{Knowledge-free Generative Recommendation}
Generative models have recently been applied to model user-item interactions. These include auto-encoding models~\cite{sedhain2015autorec,sun2019bert4rec}, which reconstruct user preference vectors from partially observed inputs; VAEs such as VAE-CF~\cite{liang2018variational} and PivotCVAE~\cite{liu2021variation}, that model latent user preference and generate complete recommendation slates; auto-regressive models~\cite{hidasi2015session} and Transformer-based architectures for sequential recommendation. Additionally, GANs~\cite{goodfellow2014generative,xu2019modeling,wang2017irgan} have been employed to sample hard negatives and augment training data, while diffusion models such as DiffRec~\cite{wang2023diffusion}, SDRM~\cite{lilienthal2024multi}, and GiffCF~\cite{zhu2024graph} remodel interactions via diffusion processes. These methods collectively form the basis of interaction-augmented generative recommendation.

\section{Methodology}
In this section, we formally define the recommendation problem and the objective of its generative augmentation. Next, we introduce the proposed NodeDiffRec framework in accordance with our injection-denoising workflow, depicted in Figure \ref{system overview}. We firstly outline key components of the node-level generation module, designed to conduct delicate node and edge-level augmentation. Then, we elaborate on how to attenuate the accompanying structural noise by components in the denoising preference distribution modeling module.

\subsection{Preliminaries}

Let $\mathcal{U}$ and $\mathcal{I}$ denote the sets of users and items, respectively, where $\mathcal{U}$ has $|\mathcal{U}| = M$ users and $\mathcal{I}$ has $|\mathcal{I}| = N$ items. The user-item interaction matrix is denoted as $\mathbf{X} \in \{0, 1\}^{M \times N}$. The goal of recommender systems is to learn a scoring function $f_\theta: \mathcal{U} \times \mathcal{I} \rightarrow \mathbb{R}$, that predicts the likelihood of potential interactions and ranks all candidates according to $f_\theta(u, i)$ for each $u \in \mathcal{U}$, to recommend users top-$K$ items:
\begin{equation}
\text{TopK}(u) = \operatorname{arg\,topK}_{i \in \mathcal{I}} f_\theta(u, i).
\label{eq:topk}
\end{equation}

To overcome data sparsity and cold-start entity challenges, generative models have been leveraged to augment the user-item interaction matrix $\mathbf{X}$. Given observed interactions $\mathbf{X} \in \mathbb{R}^{M \times N}$, a generative model $G_\phi$ seeks to learn the underlying preference distribution $\mathcal{P}$ of users $\in \mathcal{U}$ within user-item interactions over items $\in \mathcal{I}$, to reconstruct or extrapolate the interaction space. $G_\phi$ is trained to approximate $\mathcal{P}$, where $z \in \mathcal{Z}$ is sampled from a prior distribution $p(\mathbf{z})$:
\begin{equation}
\mathbf{z} \sim p(\mathbf{z}), \;\; \tilde{\mathbf{x}} = G_\phi(\mathbf{z}) \sim p_{G_\phi} \approx \mathcal{P}, \;\; G_\phi: \mathcal{Z} \rightarrow \mathbb{R}^N.
\label{eq:approximate_distribution}
\end{equation}

By repeating this process for $M'$ samples, synthetic interaction patterns $\smash{\tilde{\mathbf{X}} = \{ \tilde{\mathbf{x}}_1, \tilde{\mathbf{x}}_2, \ldots, \tilde{\mathbf{x}}_{{M'}} \}^\top}\in \mathbb{R}^{{M'} \times N}$ is obtained to augment original interactions: $\mathbf{X}^{+} = \mathcal{A}(\mathbf{X}, \smash{\tilde{\mathbf{X}}})$, where $\mathcal{A}(\cdot)$ denotes generic augmentation operators, typically to replace $\mathbf{X}$ with $\smash{\tilde{\mathbf{X}}}$ as an optimized matrix, when $M'$ equals $M$. In addition, we endeavor to synthesize new entities $\tilde{\mathcal{U}}$, $\tilde{\mathcal{I}}$, and their informative interactions $\tilde{\mathcal{E}}$, augmenting the user-item graph as $\mathcal{G}^{+} = (\mathcal{A}(\mathcal{U}, \smash{\tilde{\mathcal{U}}}) \cup \mathcal{A}(\mathcal{I}, \smash{\tilde{\mathcal{I}}}), \mathcal{A}(\mathbf{X}, \smash{\tilde{\mathcal{E}}}))$. Recommendation algorithms are then trained upon $\mathbf{X}^{+}$ or $\mathcal{G}^{+}$ to learn the scoring function $f_\theta(u, i)$ introduced in Eq.\ref{eq:topk}.

\subsection{Node-level Graph Generation}\label{Node-level Graph Generation}
\subsubsection{Embedding Initialization}
NodeDiffRec begins by pretraining a LightGCN embedder~\cite{he2020lightgcn} to construct position-aware node initialization. This extracts collaborative signals derived from interactions $\mathbf{X}$, which is treated as a bipartite graph $\mathcal{G} = (\mathcal{U} \cup \mathcal{I}, \mathcal{E})$. After $k$-layer propagation steps cross $\mathcal{G}$, the initial input embedding $z_v \in \mathbb{R}^d$ of entity node $v$ is updated by the embeddings of its neighbors \( \mathcal{N}(v) \):
\begin{equation}
z_v = \frac{1}{K+1} \sum_{k=0}^{K} \sum_{u \in \mathcal{N}(v)} \frac{1}{\sqrt{|\mathcal{N}(v)|} \sqrt{|\mathcal{N}(u)|}} \, z_u^{(k-1)}.
\end{equation}

\subsubsection{Injection Encoder}
To complement the structure-aware embeddings, we incorporate a positional embedding matrix $P \in \mathbb{R}^{(M+N) \times d}$~\cite{vaswani2017attention} to enhance topological information. The encoder first applies a linear transformation on the input $z_v$, followed by a ReLU activation $\sigma(\cdot)$:
\begin{equation}
    h_v^{(0)} = \sigma(W_f z_v + b_f), \quad \forall v \in \mathcal{V}=\mathcal{U} \cup \mathcal{I},
\end{equation}
where $W_f \in \mathbb{R}^{d_h \times d}$, $b_f \in \mathbb{R}^{d_h}$, $d_h$ is the hidden dimension.

Let $H^{(0)} = [h_1^{(0)}, \ldots, h_{M+N}^{(0)}]^\top \in \mathbb{R}^{(M+N) \times d_h}$, and define $P'$ as the projection of the positional embeddings:
\begin{equation}
    P' = W_p P + b_p, \; W_p \in \mathbb{R}^{d_h \times d}, \; H_{\text{input}} = H^{(0)} + P'.
\end{equation}

A LightGCN-based architecture with $L$ layers is exploited to aggregate neighbor information via normalized aggregation. The input to the first layer $H^{(0)}$ is set as $H_{\text{input}}$:
\begin{equation}
    H^{(\ell)} = A_{Norm} H^{(\ell-1)}, \quad \ell = 1, \ldots, L,
\end{equation}
where $A_{Norm} = \tilde{D}^{-1/2} \tilde{X} \tilde{D}^{-1/2}$ is the normalized adjacency matrix with self-loops, i.e., $\tilde{X} = X + I$, and $\tilde{D}_{ii} = \sum_j \tilde{X}_{ij}$.

$H^{(0)}$ and the neighborhood encoding $H_{\text{neigh}}= H^{(L)}$ are then concatenated to form the joint representation:
\begin{equation}
    H_{\text{joint}} = \left[ H^{(0)} \parallel H_{\text{neigh}} \right] \in \mathbb{R}^{(M+N) \times 2d_h}.
\end{equation}

Next, the concatenated embeddings pass through a feedforward layer to obtain intermediate hidden representations:
\begin{equation}
    H_z = \sigma(W_z H_{\text{joint}} + b_z), \quad W_z \in \mathbb{R}^{d_r \times 2d_h}.
\end{equation}
We denote variational parameters $\mu_v = W_\mu H_z + b_\mu$ as the \textbf{Mean} of latent Gaussian, $\log \sigma_v^2 = W_{\log \sigma} H_z + b_{\log \sigma},$ as the \textbf{Log-variance}, where $W_\mu, W_{\log \sigma} \in \mathbb{R}^{d_z \times d_r}$. Through reparameterization tricks~\cite{kingma2013auto}, we sample latent variable $z_v^{\text{latent}}$ from:
\begin{equation}
    z_v = \mu_v + \epsilon \odot \sigma_v,\; \epsilon \sim \mathcal{N}(0,I),\; \sigma_v = \exp\left(\! \tfrac{1}{2} \log \sigma_v^2 \!\right).
\end{equation}
Thus, $q_\phi(z_v^{\text{latent}} | z_v^{\text{in}}, X) = \mathcal{N}(z_v^{\text{latent}} | \mu_v, \mathrm{diag}(\sigma_v^2))$ defines the variational posterior, parameterized by neural networks.

\subsubsection{Diffusion}
We employ a conditional denoising diffusion probabilistic model (DDPM)~\cite{ho2020denoising} over $z_v^{\text{latent}}$ to generate plausible items and interactions for injection. Specifically, the forward diffusion process that gradually adds Gaussian noise $\epsilon$ to \( z_v^{\text{latent}} \) over \( T \) time steps:
\begin{equation}
    q(z_v^{(t)} | z_v^{(t-1)}) = \mathcal{N}(z_v^{(t)} ; \sqrt{1 - \beta_t} z_v^{(t-1)}, \beta_t \mathbf{I}),
\end{equation}
where \( \beta_t \in (0,1) \) is the noise variance at \( t \). Let \( \alpha_t = 1 - \beta_t \), and the cumulative product \( \bar{\alpha}_t = \prod_{s=1}^t \alpha_s \) represents the proportion of the original signal retained after noise injection. Accordingly, the marginal distribution is derived: 
\begin{equation}
    q(z_v^{(t)} | z_v^{\text{latent}}) = \mathcal{N}(z_v^{(t)} ; \sqrt{\bar{\alpha}_t} z_v^{\text{latent}}, (1 - \bar{\alpha}_t) \mathbf{I}).
\end{equation}

The reverse process can be parameterized as:
\begin{align}
p_\theta(z_v^{(t-1)} \mid z_v^{(t)}, c_v) 
&= \mathcal{N}\big(z_v^{(t-1)}; \mu_\theta(z_v^{(t)}, t, c_v),\notag\\
& \Sigma_\theta(z_v^{(t)}, t)\big), \; c_v = Embed(y_v),
\end{align}
where the conditional signal $c_v$ is obtained from the entity class label $y_v$, indicating whether the node is a user or an item. The mean $\mu_\theta(\cdot)$ is calculated by a denoising U-Net network $\epsilon_\theta(z_v^{(t)}, t, c_v)$, trained to predict the noise ${\epsilon}_\theta$ added in the forward process:
\begin{equation}
\mu_\theta(z^{(t)}, t, c) = \frac{1}{\sqrt{\alpha_t}} \left( z^{(t)} - \frac{1 - \alpha_t}{\sqrt{1 - \bar{\alpha}_t}} \cdot \epsilon_\theta(z^{(t)}, t, c) \right).
\end{equation}

During training, $\epsilon_\theta(z_v^{(t)}, t, c_v)$ is optimized by minimizing the expected error between true noise $\epsilon$ and the predicted $\epsilon_\theta$:
\begin{equation}
\mathcal{L}_{\text{diff}} = \mathbb{E}_{z_v, \epsilon, t} \left[ \left\| \epsilon - \epsilon_\theta(z_v^{(t)}, t, c_v) \right\|^2 \right].
\end{equation}

\subsubsection{Injection Decoder}
The decoder reconstructs both node features and the user-item interaction map from the latent vector $z \in \mathbb{R}^{d_z}$, which is sampled from the variational posterior. First, $z$ is transformed and split into $z_f'$ and $z_a'$, where $z_f'$ is used to construct features of newly generated items, and $z_a'$ is used to predict the user-item interaction map:
\begin{equation}
z' = \text{ReLU}(W_2 \cdot \text{ReLU}(W_1 \cdot z + b_1) + b_2) = [z_f' \, \| \, z_a'].
\end{equation}

The reconstructed node features $\hat{\mathbf{Z}} \in \mathbb{R}^{(M+N) \times d_f}$ and the predicted user-item interaction map $\hat{\mathbf{X}}$ are obtained by:
\begin{equation}
\hat{Z} = \sigma(W_f \cdot z_f' + b_f), \;
\hat{X} = \sigma\left( W_a \cdot z_a' + b_a \right).
\end{equation}
Only the user-item block of the predicted $\hat{X}$ is retained.

The injection encoder and decoder are trained jointly to minimize the reconstruction error for both node features and the user-item adjacency map, combined as:
\begin{equation}
\mathcal{L}_{\text{total}} = \lambda_{\text{feat}} \cdot \| \hat{Z} - Z \|_2^2 + \lambda_{\text{map}} \cdot \text{BCE}(\hat{X}, X).
\end{equation}

\subsection{User-Item Graph Enrichment}
We sample latent vectors $\{ z^{(i)}_{\text{new}} \}_{i=1}^{N'}$ from the learned latent space, which are then decoded by the injection decoder to obtain node features and interaction patterns of new items, that connect them to existing users:
\begin{equation}
\hat{\mathbf{Z}}_{\text{new}} = \sigma(W_f \cdot z_f^{\prime} + b_f), \quad
\hat{\mathbf{X}}_{\text{new}} = \sigma(W_a \cdot z_a^{\prime} + b_a).
\end{equation}

A threshold $\tau$ to the predicted interaction scores is applied, resulting in a sparse matrix $\tilde{\mathbf{X}}_{\text{new}} \in \mathbb{R}^{M \times N'}$ that retains only high-confidence user-item connections.

To control the volume of augmentation and reduce noise, $K$ high-confidence user-item pairs $\mathbf{X}'$ are selected from $\tilde{\mathbf{X}}_{\text{new}}$ to augment the original user-item matrix $\mathbf{X}$ by entity injection $\mathbf{X}_{\text{aug}} = [\mathbf{X} \;|\; \mathbf{X}'] \in \mathbb{R}^{M \times (N + N')}$.

\subsection{Denoising Preference Distribution Modeling}
\subsubsection{Preference Modeling VAE}
As we augment the bipartite graph $\mathcal{G}$ through graph generation, a great deal of structural noise is inevitably introduced into $\mathbf{X}_{\text{aug}}$, which may directly degrade the performance of downstream recommendations. In response, we employ another score-based diffusion to remodel users' preferences, aiming to eliminate maintained unclear patterns and structural noise.

A MLP preference distribution encoder $f_{\text{enc}}$ first maps $\mathbf{X}_{\text{aug}}$ into a latent representation $\mathbf{z} \in \mathbb{R}^d$ and outputs the mean $\boldsymbol{\mu} \in \mathbb{R}^d$ and log-variance $\log \boldsymbol{\sigma}^2 \in \mathbb{R}^d$. The latent vector $\mathbf{z}$ is obtained using the parameterization trick:
\begin{equation}
    \mathbf{z} = \boldsymbol{\mu} + \boldsymbol{\sigma} \odot \boldsymbol{\epsilon}, \quad \boldsymbol{\epsilon} \sim \mathcal{N}(0, \mathbf{I}).
\end{equation}

The denoising preference reconstructor $f_{\text{dec}}$ reconstructs optimized interactions $\mathbf{X}_{\text{opt}} = f_{\text{dec}}(\mathbf{z})$, and the VAE is trained to minimize the negative evidence lower bound:
\begin{equation}
    \mathcal{L} = \mathbb{E}_{q(\mathbf{z}|\mathbf{x})}[-\log p(\mathbf{x}|\mathbf{z})] + \mathrm{KL}(q(\mathbf{z}|\mathbf{x}) \parallel \mathcal{N}(0, \mathbf{I})).
\end{equation}

\subsubsection{Latent Diffusion}
The VAE is frozen after training, and used to encode users' interaction behavior into latent vectors $\mathbf{z}_0$. We then train a DDPM in this latent space, where the forward process gradually perturbs each latent vector by adding Gaussian noise over \( T \) steps following a linear variance schedule \( \{\beta_t\}_{t=1}^T \) with cumulative product \( \bar{\alpha}_t \):
\begin{equation}
    \mathbf{z}_t = \sqrt{\bar{\alpha}_t} \cdot \mathbf{z}_0 + \sqrt{1 - \bar{\alpha}_t} \cdot \boldsymbol{\epsilon}, \quad \boldsymbol{\epsilon} \sim \mathcal{N}(0, \mathbf{I}).
\end{equation}

A denoising network \( \epsilon_\theta(\mathbf{z}_t, t) \) is trained subsequently to estimate the added noise \( \boldsymbol{\epsilon} \), which is equivalent to learning the negative score function \( -\nabla_{\mathbf{z}_t} \log p(\mathbf{z}_t | t) \). Instead of computing the score directly, we adopt a finite-difference approximation to train \( \epsilon_\theta \). The training objective becomes:
\begin{equation}
\begin{aligned}
\mathcal{L}_{\text{diff}} = \frac{1}{2} \cdot \frac{1}{\mathrm{Var}[\boldsymbol{\epsilon} - \mathbf{z}_t]} \cdot \bigg(
\left\| \frac{s_\theta(\mathbf{z}_t + \mu \cdot \boldsymbol{\epsilon}, t)}{\mu^2} 
\right. \\
\left.
- \frac{s_\theta(\mathbf{z}_t, t)}{\mu^2} - (\boldsymbol{\epsilon} - \mathbf{z}_t) \right\|^2
+ \left\| s_\theta(\mathbf{z}_t, t) - (\boldsymbol{\epsilon} - \mathbf{z}_t) \right\|^2 \bigg).
\end{aligned}
\end{equation}

\subsubsection{Sampling}
At last, latent vectors \( \mathbf{z}_T \sim \mathcal{N}(0, \mathbf{I}) \) are sampled by a reverse diffusion process, which iteratively refines this vector with the trained \( \epsilon_\theta(\mathbf{z}_t, t) \). After \( T \) steps, the denoised \( \mathbf{z}_0 \) is decoded by the pretrained preference modeling VAE to produce the final optimized interactions $\mathbf{X}_{\text{opt}}$, which is utilized to run recommendation algorithms thereafter.

\begin{table*}[t]
\centering
\caption{Performance comparison of recommendation algorithms before and after NodeDiffRec (*) across datasets.}
\label{tab:Recall_NDCG_comparison}
\resizebox{\textwidth}{!}{
\renewcommand{\arraystretch}{2.5}
\begin{tabular}{l|cccc|cccc|cccc|cccc|cccc|cccc}
\toprule
& \multicolumn{8}{c|}{\huge\textbf{ProgrammableWeb}} 
& \multicolumn{8}{c|}{\huge\textbf{Amazon Luxury Beauty}} 
& \multicolumn{8}{c}{\huge\textbf{MovieLens 100K}} \\
\cmidrule(r){2-9} \cmidrule(r){10-17} \cmidrule(r){18-25}
\huge{Algorithm} 
& \multicolumn{4}{c|}{\huge Recall} & \multicolumn{4}{c|}{\huge NDCG}
& \multicolumn{4}{c|}{\huge Recall} & \multicolumn{4}{c|}{\huge NDCG}
& \multicolumn{4}{c|}{\huge Recall} & \multicolumn{4}{c}{\huge NDCG} \\
\cmidrule(r){2-5} \cmidrule(r){6-9}
\cmidrule(r){10-13} \cmidrule(r){14-17}
\cmidrule(r){18-21} \cmidrule(r){22-25}
& {\huge @5}  & {\huge @10}  & {\huge @20}  & {\huge @50}
& {\huge @5}  & {\huge @10}  & {\huge @20}  & {\huge @50}
& {\huge @5}  & {\huge @10}  & {\huge @20}  & {\huge @50}
& {\huge @5}  & {\huge @10}  & {\huge @20}  & {\huge @50}
& {\huge @5}  & {\huge @10}  & {\huge @20}  & {\huge @50}
& {\huge @5}  & {\huge @10}  & {\huge @20}  & {\huge @50} \\
\bottomrule
\end{tabular}
}
\end{table*}

\section{Experiment}
In this section, we study the following research questions:
\begin{itemize}
\item \textbf{RQ1}: Can NodeDiffRec improve recommendation performance across a wide range of algorithms and datasets?
\item \textbf{RQ2}: How does NodeDiffRec compare to existing generative baselines (SOTA) in terms of generation quality?
\item \textbf{RQ3}: How each component, especially the node-level graph generation, contributes to the performance gain.
\item \textbf{RQ4}: In what manner does NodeDiffRec reshape the user-item interactions and benefit recommendation?
\end{itemize}

\subsection{Experimental Setup} \label{Experimental Setup}
\subsubsection{Datasets}
We conduct experiments on three public benchmark datasets from multiple domains, including \textbf{Amazon Luxury Beauty}, \textbf{ProgrammableWeb}, and \textbf{MovieLens-100k}. More details are available in the Appendix Section~\ref{Statistics of Dataset}.

\subsubsection{Algorithms \& Baselines} NodeDiffRec is evaluated across a diverse set of representative recommendation algorithms introduced in Section~\ref{sec:Recommendation Algorithm}, well-performing generative models, and SOTA DMs for recommendation. 

Recommendation algorithms include: 1) \textbf{ALS}~\cite{hu2008collaborative}; 2) \textbf{DR-MF}~\cite{halko2011finding}, a dimensionality reduction matrix factorization implemented via TruncatedSVD; 3) \textbf{PureSVD}~\cite{cremonesi2010performance}; 4) \textbf{NGCF}~\cite{wang2019neural}; 5) \textbf{NeuMF}~\cite{he2017neural}; 6) \textbf{NGSGL}, a neural graph model with self-supervised signals from SGL~\cite{wu2021self}; 7) \textbf{NGSimGCL}, a neural graph model with contrastive signals from SimGCL~\cite{yu2022graph}; and 8) \textbf{HybridRec}, a hybrid model mixing ALS and SVD.

For comparison, we consider the following generative baselines: 9) \textbf{MultiVAE} from VAE-CF~\cite{liang2018variational}, a VAE-based model modeling implicit feedback as probabilistic generative processes; 10) \textbf{TVAE}~\cite{xu2019modeling}, a conditional tabular VAE tailored for synthetic tabular data generation; 11) \textbf{VanillaGAN}~\cite{goodfellow2014generative}, a Vanilla GAN with MLP-based generator and discriminator, to generate user-item interactions; 12) \textbf{CTGAN}~\cite{xu2019modeling}, a conditional GAN designed to generate realistic synthetic tabular data; 13) \textbf{DiffRec}, a DM learning the denoising generation process for personalized recommendation, becoming SOTA at the time; 14) \textbf{GiffCF}, a DM that replaces standard Gaussian corruption with a graph signal smoothing process, outperforming DiffRec to become the current SOTA; 15) \textbf{SDRM}, a DM that learns interaction representations through a multi-resolution reverse diffusion process, emerging as another SOTA. F-SDRM and M-SDRM are its two variations with full-time and multi-resolution sampling, respectively.

\subsubsection{Implementation} For the injection VAE, the learning rate (lr) and training epochs are set to 0.0002 and 30000. Generated items number ${N'}$ is 2000, and the confidence threshold $\tau$ is set to 1.0. The number of high-confidence interactions $K$ is tuned at intervals of 500 for different algorithms. For denoising preference distribution modeling, we search hyperparameters in terms of Recall@10: diffusion training epochs in \{5, ..., 500\} with step 5, diffusion lr in \{1e-6, ..., 1e-4\} with step 1e-6, denoising timesteps in \{3, ..., 200\} with step 5, VAE latent dimension in \{20, ..., 1000\} with step 10, and VAE lr in \{1e-4, ..., 1e-2\} with step 1e-4.

\subsection{Main Results}
\subsubsection{Recommendation Performance (Q1)}
To quantize the performance boost achieved by NodeDiffRec, we conduct experiments on all eight recommendation algorithms listed in Section~\ref{Experimental Setup}, and report standard top-K Recall and NDCG. As presented in Table~\ref{tab:Recall_NDCG_comparison}, NodeDiffRec consistently yields performance across all settings with a maximum 69.95\% improvement in Recall and 57.14\% in NDCG. Particularly, our augmentation leads to almost equal or more than 40\% performance gain with all algorithms on the ProgrammableWeb dataset, where the improvement in Recall is more pronounced. This might be due to the objective metric of hyperparameter search, Recall@10. The improvement is significant on the ML-100K dataset as well, with a maximum 57.45\% improvement in Recall and 57.34\% in NDCG. In addition, non-trivial performance gain is also observed on the ALB dataset, with the maximum 26.22\% Recall and 40.43\% NDCG improvement. The improvements are especially notable on the ProgrammableWeb dataset, with most 99.73\% extreme data sparsity, demonstrating the necessity and superiority of  NodeDiffRec in low-resource scenarios.

Strikingly, NodeDiffRec achieves statistically significant improvements against substantial variation in the underlying interaction characteristics and distribution across datasets, modeling assumption and mechanism variation of recommendation algorithms. This underscores that our framework provides robust augmentation applicable to real-world recommendation scenarios and exhibits strong generalization.

\subsubsection{Comparison Against Generative Baselines (Q2)}

\begin{figure}[h]
\centering
\includegraphics[width=\linewidth,height=7cm]{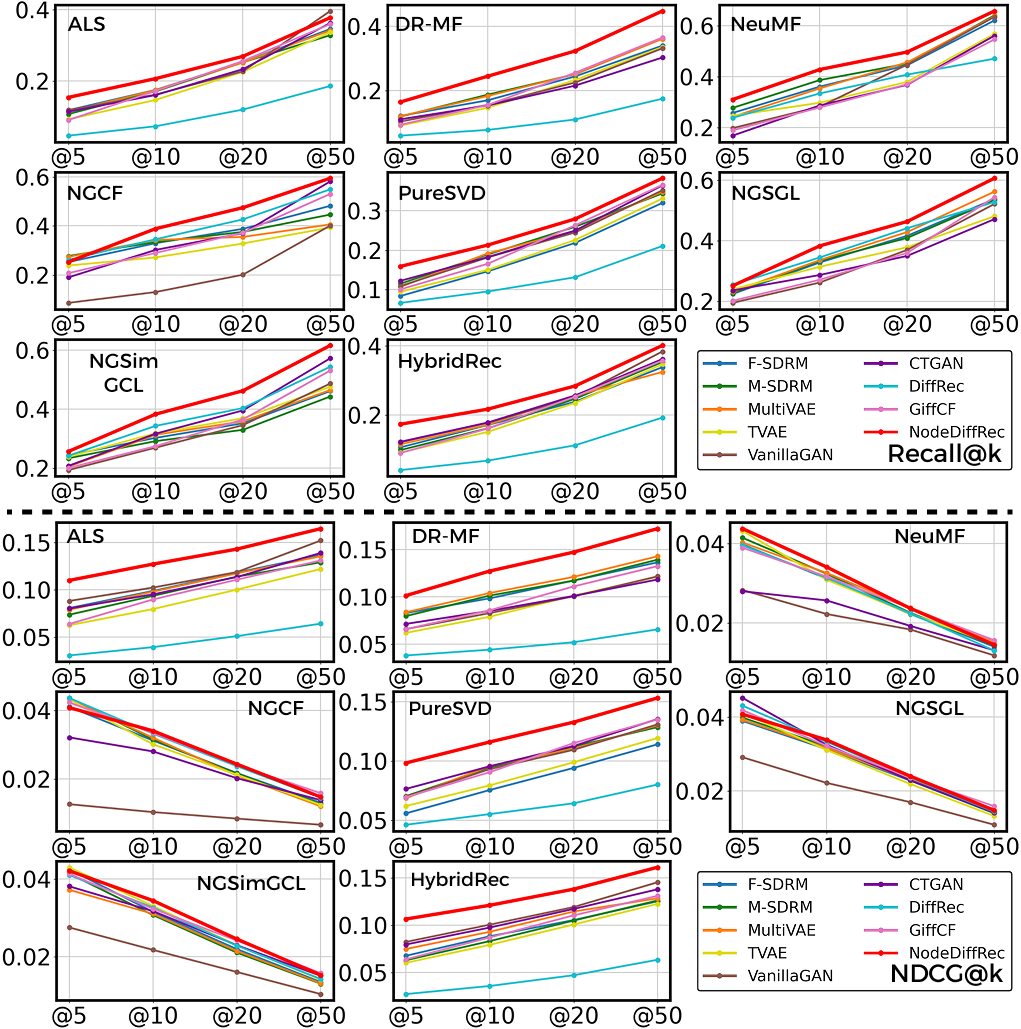}
\caption{Comparison against generative baselines on ProgrammableWeb. (Upper: Recall@k, Lower: NDCG@k.)}
\label{Generative baselines}
\end{figure}

\begin{table}[h]
\centering
\caption{Average relative improvement ratio over generative baselines on ProgrammableWeb dataset.}
\label{tab:Average relative improvement}
\renewcommand{\arraystretch}{2}
\resizebox{\linewidth}{!}{%
\begin{tabular}{lcccccccc}
\toprule
& 
\multicolumn{4}{c}{\textbf{\Huge{Recall}}} & 
\multicolumn{4}{c}{\textbf{\Huge{NDCG}}} \\
\cmidrule(lr){2-5} \cmidrule(lr){6-9}
\makecell{\textbf{\Huge{Algorithm}}} & 
\makecell[c]{\Huge@5} & \makecell[c]{\Huge@10} & \makecell[c]{\Huge@20} & \makecell[c]{\Huge@50} 
& \makecell[c]{\Huge@5} & \makecell[c]{\Huge@10} & \makecell[c]{\Huge@20} & \makecell[c]{\Huge@50} \\
\midrule
\Huge{ALS} & 
\Huge 68.0\% & \Huge45.0\% & \Huge25.5\% & \Huge19.4\% & 
\Huge74.8\% & \Huge59.9\% & \Huge46.1\% & \Huge39.3\% \\

\Huge DR-MF & 
\Huge 69.0\% & \Huge\textbf{70.9\%} & \Huge\textbf{55.5\%} & \Huge\textbf{46.9\%} & 
\Huge56.3\% & \Huge59.8\% & \Huge\textbf{53.3\%} & \Huge\textbf{49.0\%} \\

\Huge NeuMF & 
\Huge 40.2\% & \Huge34.6\% & \Huge20.5\% & \Huge13.6\% & 
\Huge19.9\% & \Huge16.4\% & \Huge9.1\% & \Huge4.6\% \\

\Huge NGCF & 
\Huge 29.7\% & \Huge45.6\% & \Huge41.3\% & \Huge28.1\% & 
\Huge27.6\% & \Huge36.2\% & \Huge33.2\% & \Huge24.0\% \\

\Huge PureSVD & 
\Huge 64.6\% & \Huge37.1\% & \Huge26.9\% & \Huge19.5\% & 
\Huge55.1\% & \Huge40.9\% & \Huge34.2\% & \Huge29.2\% \\

\Huge HybridRec & 
\Huge \textbf{98.6\%} & \Huge53.9\% & \Huge32.0\% & \Huge26.0\% & 
\textbf{\Huge84.0\%} & \Huge\textbf{61.3\%} & \Huge46.7\% & \Huge40.0\% \\

\Huge NGSGL & 
\Huge 11.8\% & \Huge24.9\% & \Huge18.3\% & \Huge16.4\% & 
\Huge4.0\% & \Huge11.5\% & \Huge9.0\% & \Huge7.6\% \\

\Huge NGSimGCL & 
\Huge 17.3\% & \Huge26.9\% & \Huge27.3\% & \Huge24.8\% & 
\Huge10.0\% & \Huge14.4\% & \Huge14.9\% & \Huge13.5\% \\

\bottomrule
\end{tabular}
}
\end{table}

\begin{figure}[h]
\centering
\includegraphics[width=8cm,height=6.5cm]{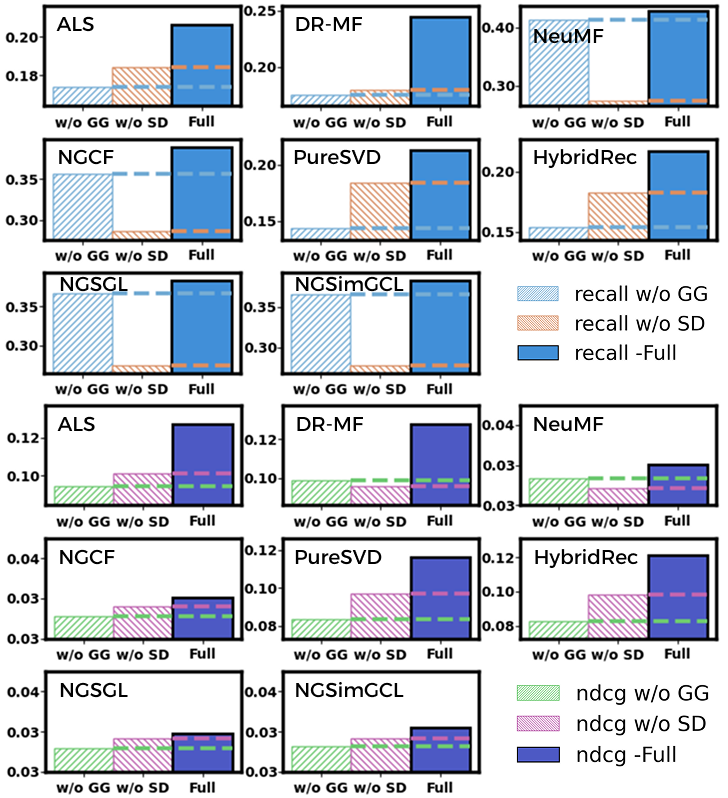}
\caption{Ablation study on ProgrammableWeb @k = 10.}
\label{Abaltion figure}
\end{figure}

Figure \ref{Generative baselines} outlines the augmentation performance of all generative baselines with selected recommendation algorithms, in which DiffRec, GiffCF, and SDRM are the previous and two current SOTA diffusion models. The complete baseline performance on all datasets is presented in Appendix Section~\ref{Complete Performance of Generative Baselines}. Across all datasets and algorithms, our NodeDiffRec demonstrates superior performance over baselines in general, in terms of both Recall and NDCG. According to Table~\ref{tab:Average relative improvement} that shows average relative improvement ratios over baselines, in terms of Recall@5, the highest improvement is achieved on HybridRec, reaching 98.6\%, followed by DR-MF 69.0\% and ALS 68.0\%. Improvements remain discernible even at larger cutoffs, with DR-MF showing the strongest performance at Recall@10 70.9\%, Recall@20 55.5\%, and Recall@50 46.9\%. Regarding NDCG, NodeDiffRec again achieves remarkable boosts. For NDCG@5, the highest relative improvement is on HybridRec 84.0\%, significantly outperforming other models. Furthermore, notable performance gains also exist across other recommendation algorithms such as ALS, NGCF, PureSVD, and NGSimGCL, ranging from 10\%+ to 40\%+.

While GiffCF, F-SDRM, and M-SDRM show strong enhancing performance in the majority of settings, NodeDiffRec achieves better augmenting capability compared with them. The consistent advantage highlights the effectiveness of our generative design and establishes our approach as a SOTA solution for recommendation augmentation.

\subsubsection{Ablation Study (Q3)}
To assess the contribution of individual components, we conduct a comprehensive ablation study on the two core modules: the node-level graph generation component, GG, and the structural denoising preference modeling component, SD. Figure~\ref{Abaltion figure} shows the ablation performance in terms of Recall@10 and NDCG@10. The results demonstrate that each component plays a crucial role. Removing GG consistently leads to noticeable performance drops across all algorithms, emphasizing the necessity of nuanced node generations and the proposed entity-injection primitive with corresponding edge augmentation for effective recommendation augmentation. Similarly, the degradation made when removing SD highlights the effectiveness of our denoising preference distribution modeling design.

\begin{figure}[t]
\centering
\includegraphics[width=7.8cm,height=4cm]{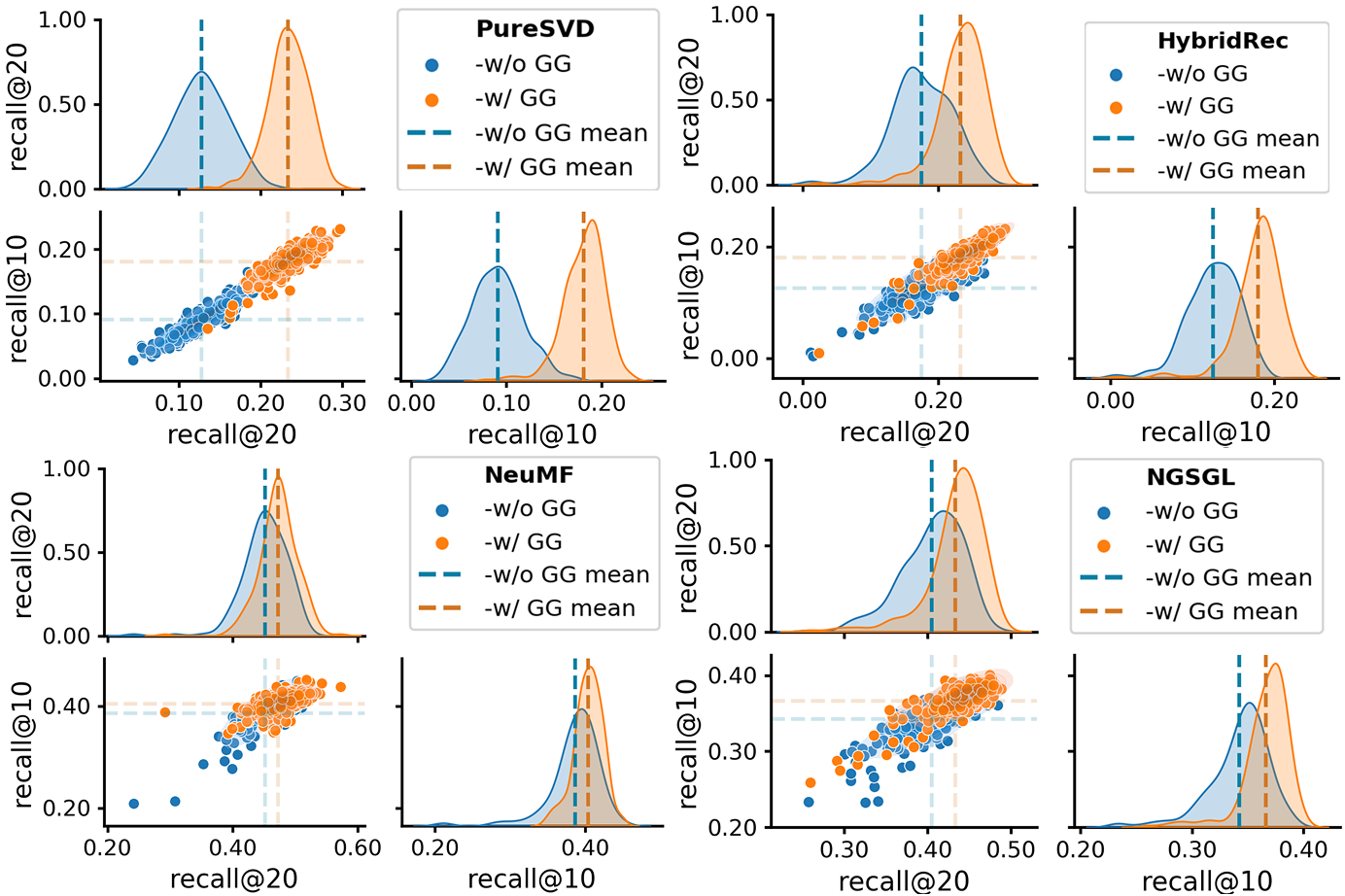}
\caption{Impact of node-level graph generation.}
\label{distribution plot}
\end{figure}

Figure~\ref{distribution plot} further confirms the importance of the node-level generation and injection operation, showing clear upper-right shifts in the paired Recall@10 and Recall@20 distributions when GG is present. The improvement across algorithms highlights GG as a key contributor to accurate recommendations. Full experimental results of the ablation study can be found in Appendix Section~\ref{Full Ablation}.

\begin{figure}[h]
\centering
\includegraphics[width=\linewidth,height=5cm]{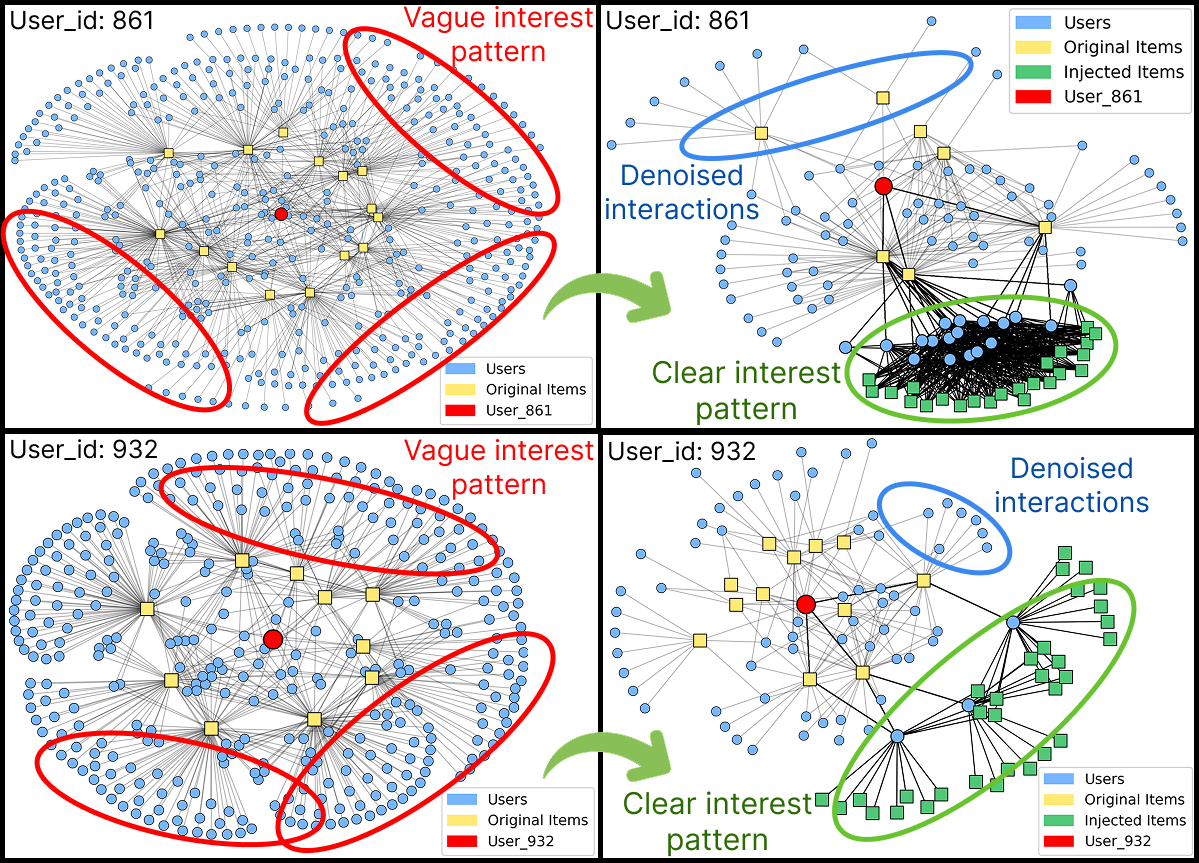}
\caption{Visualizations before and after NodeDiffRec.}
\label{case study}
\end{figure}

\subsubsection{Case Study (Q4)}
To investigate how NodeDiffRec benefits recommendation, we visualize user-item interactions before and after augmentation for two representative users. As shown in Figure~\ref{case study}, the injected items significantly increase local connectivity around target users, while the denoising preference modeling component successfully removes structural noisy interaction and vague patterns. According to Appendix Table~\ref{metrics for case study}, for User\_861, its Recall@5 is improved from 0.20 to 0.60 and NDCG@5 from 0.214 to 0.684. The Recall@50 of User\_932 increases from 0.50 to 1.0 after augmentation. These results point to the fact that NodeDiffRec effectively enhances recommendations by enriching relevant items and interactions, and exposing users' real interests.

\section{Conclusion and Future Work}
In this work, we present NodeDiffRec, the first diffusion-based knowledge-free augmentation framework that introduces node-level graph generation to enhance recommendation performance. By generating new items with informative user-item interactions and refining user preferences via denoising diffusion, our method enables fine-grained augmentation without relying on external knowledge and achieves SOTA augmentation performance. NodeDiffRec points out the significance of granular generation and the necessity of noise suppression after injection, offering practical insights for future works. We plan to extend node-level augmentation to the user side and explore more controllable generation mechanisms for recommendation tasks.

\bibliography{aaai2026}

\cleardoublepage
\appendix

\begin{center}
    \LARGE\bfseries Appendix
\end{center}
\section{Use case Analysis} \label{Use case Analysis}

\begin{figure*}[t]
\centering
\includegraphics[width=12cm]{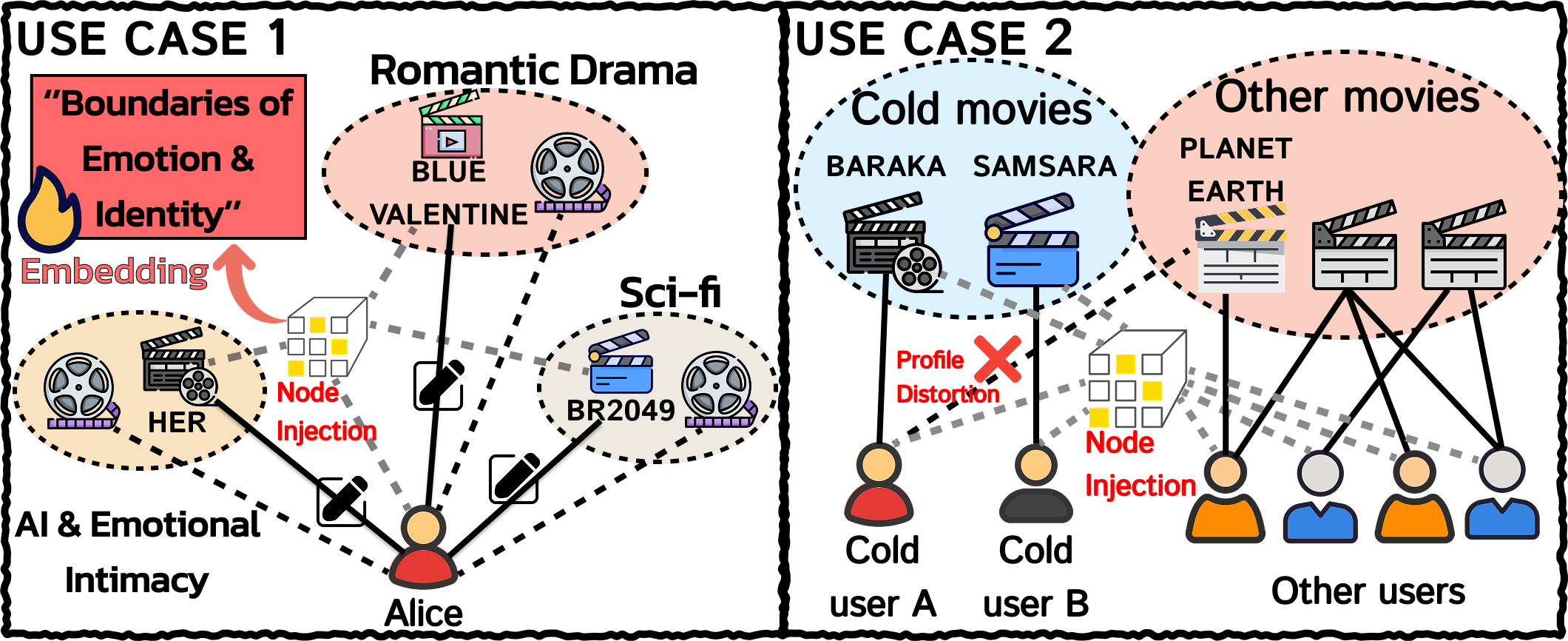}
\caption{Constraints of SOTA knowledge-free model-based augmentation: (1) Interest Expression Gaps in Embedding Space; (2) Structural Instability in Isolated Cold Entities.}
\label{use cases appendix}
\end{figure*}

Figure~\ref{use cases appendix} exhibits two use case scenarios that we mainly consider to illustrate the insufficiency and restrictions of current recommendation augmentation operations, which are user-item interaction deletion, creation, and reweighting, applied by existing diffusion-based generators. It also helps to understand the unique value of our proposed entity-injection and corresponding injection-centered edge augmentation primitives via diffusion-based node-level generation.

\subsection{Use case 1: Representing Complex User Preferences via Pseudo-nodes}
\textbf{Scenario.}  
Consider a user Alice who has rated the following movies:

\begin{itemize}
  \item \textit{Her}: explores artificial intelligence and emotional intimacy.
  \item \textit{Blue Valentine}: a raw depiction of romantic breakdown.
  \item \textit{Blade Runner 2049}: a dystopian sci-fi with philosophical undertones.
\end{itemize}

Alice appears to favor emotionally intense films, especially those with themes of futuristic relationships and psychological complexity. The phrase \textbf{\textit{Boundaries of Emotion \& Identity}} appropriately concludes Alice's interest pattern. However, in the dataset, there is no single item that jointly captures this blend of genres and emotional narratives.

\textbf{Problem with current interaction optimization approaches.}  
Traditional methods can only reinforce observed interactions (e.g., increasing edge weight between Alice and \textit{Her}), delete existing interactions that may lead to interest pattern shift, or create new links to similar items (e.g., pseudo-interaction to \textit{Ex Machina}). But these approaches operate within the space of existing items. No method can represent a new conceptual combination or semantic interpolation in the item space unless a corresponding item exists.

\textbf{Effect of entity injection.}  
Our method may sample and generate a new pseudo-item $v_{\text{emotionAI}}$ whose embedding lies between Alice's interacted items and better aligns to \textit{Boundaries of Emotion \& Identity}. This item is then connected to \textit{Her}, \textit{Blue Valentine}, and \textit{Blade Runner} indirectly through learned data distribution. Although not used for the final recommendation, this node structurally anchors a region of the graph representing Alice’s actual interest space.

\textbf{Result.}  
Alice’s embedding can now more accurately reflect her unique preference. The surrounding items also benefit from improved mutual context, leading to better collaborative filtering performance.

\textbf{Why interaction optimization fails.}  
No edge reweighting or pseudo-links can create a latent region in the embedding space that is unoccupied. Pseudo-nodes directly reshape the topology to include such latent concepts.

\subsection{Use case 2: Stabilizing Cold Entities through Structural Augmentation}
Cold entities are always a challenging difficulty that recommender systems need to solve, and commonly exist in all sorts of recommendation datasets.

\textbf{Scenario.}  
Three cold items: \textit{Baraka}, \textit{Samsara}, and \textit{Chronos}, are each rated by a single cold user:
\begin{align*}
\text{User A} \rightarrow \textit{Baraka}, \\
\text{User B} \rightarrow \textit{Samsara}, \\
\text{User C} \rightarrow \textit{Chronos}.
\end{align*}

These items and users are completely disconnected from the rest of the graph and from each other.

Each cold item is updated only via one user’s feedback, and each user only interacts with one item. The result is: (a) No mutual context or neighbor sharing. (b) No multi-hop structure to generalize similarities. (C) Highly unstable embeddings during training, often collapsing to default directions.

\textbf{Problem with current interaction optimization approaches.}
Edge reweighting and deletion cannot help because there is only one edge per item. Creating pseudo-interactions across unrelated cold users risks injecting noise and distorting user profiles.

\textbf{Effect of entity injection.} 
A pseudo-item node $v_{\text{docu\_link}}$ is sampled by the diffusion model in the embedding space between the three cold items. This node is connected to all three cold items, and optionally to a known item like \textit{Planet Earth}. This results in the following structure:
\begin{align*}
\text{A} &\rightarrow \textit{Baraka} \leftrightarrow v_{\text{docu\_link}} \leftrightarrow \textit{Samsara} \leftarrow \text{B}, \\
        &\quad\quad\quad\quad v_{\text{docu\_link}} \rightarrow \textit{Planet Earth}.
\end{align*}

\textbf{Result.}  
Cold items are now part of a connected subgraph. During training, their embeddings benefit from indirect updates propagated through $v_{\text{docu\_link}}$, leading to more stable and semantically grounded representations.

\textbf{Why interaction optimization fails.}  
There is no edge-level operation that can introduce a structural bottleneck or path between cold regions without distorting user profiles. Only an injected new node with learned structural connectivity can enable this kind of stabilization.

\subsubsection{Summary}
In summary, the above cases demonstrate that current primitives of diffusion-based generators fall short in bridging semantic and structural gaps as high-quality paradigm substitutes, motivating the necessity for expressive and topology-aware augmentation primitive expansion. Node injection by generating plausible new nodes with corresponding interactions addresses the challenges in both use cases directly: it enables semantic coverage of latent user interests via synthetic nodes, and provides structurally meaningful bridges for cold entities.

\section{Statistics of Datasets} \label{Statistics of Dataset}
Table~\ref{tab:dataset-stats} shows details of the datasets used in our experiments. We evaluate NodeDiffRec on three benchmark datasets:
\begin{itemize}
\item \textbf{ProgrammableWeb (ProgWeb)}: This dataset is constructed based on data crawled from ProgrammableWeb, a well-known directory of web APIs and mashups. Each mashup typically integrates multiple APIs to achieve a specific functionality. In our setting, the recommendation task involves predicting suitable APIs (items) for a given mashup (user), making it a form of item recommendation in a service composition context, consistent with prior works on API or service recommendation.
\item \textbf{Amazon Luxury Beauty (ALB)}: The ALB dataset is a subset of the Amazon product review corpus, specifically focusing on the Luxury Beauty category. It contains user purchase and review behavior, where user-item interactions are derived from review history.
\item \textbf{MovieLens-100k (ML-100K)}: ML-100K is a classical benchmark dataset widely used in collaborative filtering research. It contains explicit user ratings on movies, with each interaction labeled with a rating score. To make recommendations more challenging, we randomly drop partial interactions to render higher sparsity.
\end{itemize}

DiffRec and GiffCF provide recommendations as ranked lists with associated weight scores. However, they do not employ a deterministic algorithm for edge generation. Since the generative models do not guarantee a consistent range of weight scores, applying a fixed threshold directly to the raw scores is not appropriate. Consequently, we first performed min-max normalization on the output matrices to compute relative confidence scores. After normalization, we selected the edges to be generated by applying a fixed threshold, which we set to 0.5 in our experiments.

\begin{table}[h]
\centering
\caption{Statistics of the datasets used in experiments.}
\resizebox{\columnwidth}{!}{%
\begin{tabular}{ccccc}
\hline
\textbf{Dataset} & \textbf{\#Users} & \textbf{\#Items} & \textbf{\#Interactions} & \textbf{\#Sparsity} \\
\hline
ProgWeb & 2778 & 1207 & 9171 & 99.73\% \\
ALB & 1344 & 729 & 15359 & 98.43\% \\
ML-100K & 938 & 1008 & 20312 & 97.85\% \\
\hline
\end{tabular}%
}
\label{tab:dataset-stats}
\end{table}

\section{Complete Augmentation Performance of Generative Baselines} \label{Complete Performance of Generative Baselines}

In this section, complete experimental results for performance comparison with generative baselines on all three recommendation datasets are presented. Involved generative baselines include: 1). VAE-CF (Multi-VAE), 2). TVAE, 3). Vanilla-GAN, 4). CTGAN, 5). DiffRec (previous SOAT), 6). GiffCF (current SOTA), 7). F-SDRM (current SOTA), and 8). M-SDRM (current SOTA). These baselines are compared against the proposed NodeDiffRec (Ours).

For each baseline, we evaluate its augmentation performance across diverse recommendation algorithms, including: 1). ALS, 2). DR-MF, 3). PureSVD, 4). NGCF, 5). NeuMF, 6). NGSGL, 7). NGSimGCL, and 8). HybridRec.

Experimental results are shown in Table~\ref{tab:ALS} $\sim$ Table~\ref{tab:HybridRec}.

\begin{table*}[htbp]
\centering
\caption{Performance comparison against generative baselines on \textbf{ALS} recommendation algorithm.}
\resizebox{\textwidth}{!}{
\renewcommand{\arraystretch}{2.5}
\begin{tabular}{lcccccccccccccccccccccccc}
\toprule
\multicolumn{8}{c}{\hspace{15em}\huge ProgrammableWeb} &
\multicolumn{8}{c}{\hspace{15em}\huge Amazon Luxury Beauty} & 
\multicolumn{8}{c}{\hspace{15em}\huge MovieLens 100K} \\
\cmidrule(r){2-9} \cmidrule(r){10-17} \cmidrule(r){18-25}
\textbf{\huge Method} & \multicolumn{4}{c}{\huge Recall} & \multicolumn{4}{c}{\huge NDCG} 
& \multicolumn{4}{c}{\huge Recall} & \multicolumn{4}{c}{\huge NDCG} 
& \multicolumn{4}{c}{\huge Recall} & \multicolumn{4}{c}{\huge NDCG}\\
& \huge@5 & \huge@10 & \huge@20 & \huge@50 & \huge@5 & \huge@10 & \huge@20 & \huge@50
& \huge@5 & \huge@10 & \huge@20 & \huge@50 & \huge@5 & \huge@10 & \huge@20 & \huge@50
& \huge@5 & \huge@10 & \huge@20 & \huge@50 & \huge@5 & \huge@10 & \huge@20 & \huge@50 \\
\midrule
\huge F-SDRM       & \makecell{\LARGE ${0.1175}$ \\ $\pm 0.0100$} & \makecell{\LARGE ${0.1745}$ \\ $\pm 0.0051$} & \makecell{\LARGE ${0.2518}$ \\ $\pm 0.0198$} & \makecell{\LARGE ${0.3453}$ \\ $\pm 0.0088$} & \makecell{\LARGE ${0.0806}$ \\ $\pm 0.0060$} & \makecell{\LARGE ${0.0988}$ \\ $\pm 0.0044$} & \makecell{\LARGE ${0.1183}$ \\ $\pm 0.0051$} & \makecell{\LARGE ${0.1369}$ \\ $\pm 0.0043$}& \makecell{\LARGE ${0.3001}$ \\ $\pm 0.0141$} & \makecell{\LARGE ${0.3216}$ \\ $\pm 0.0130$} & \makecell{\LARGE ${0.3470}$ \\ $\pm 0.0142$} & \makecell{\LARGE ${0.3832}$ \\ $\pm 0.0140$}& \makecell{\LARGE ${0.2918}$ \\ $\pm 0.0160$} & \makecell{\LARGE ${0.3006}$ \\ $\pm 0.0147$} & \makecell{\LARGE ${0.3087}$ \\ $\pm 0.0144$} & \makecell{\LARGE ${0.3194}$ \\ $\pm 0.0138$}& \makecell{\LARGE ${0.2993}$ \\ $\pm 0.0184$} & \makecell{\LARGE ${0.2849}$ \\ $\pm 0.0090$} & \makecell{\LARGE ${0.3032}$ \\ $\pm 0.0104$} & \makecell{\LARGE ${0.3977}$ \\ $\pm 0.0131$}& \makecell{\LARGE ${0.3186}$ \\ $\pm 0.0267$} & \makecell{\LARGE ${0.3008}$ \\ $\pm 0.0173$} & \makecell{\LARGE ${0.2977}$ \\ $\pm 0.0162$} & \makecell{\LARGE ${0.3305}$ \\ $\pm 0.0152$}\\
\bottomrule
\end{tabular}
}
\label{tab:ALS}
\end{table*}

\begin{table*}[htbp]
\centering
\caption{Performance comparison against generative baselines on \textbf{DR-MF} recommendation algorithm.}
\resizebox{\textwidth}{!}{
\renewcommand{\arraystretch}{2.5}
\begin{tabular}{lcccccccccccccccccccccccc}
\toprule
\multicolumn{8}{c}{\hspace{15em}\huge ProgrammableWeb} &
\multicolumn{8}{c}{\hspace{15em}\huge Amazon Luxury Beauty} & 
\multicolumn{8}{c}{\hspace{15em}\huge MovieLens 100K} \\
\cmidrule(r){2-9} \cmidrule(r){10-17} \cmidrule(r){18-25}
\textbf{\huge Method} & \multicolumn{4}{c}{\huge Recall} & \multicolumn{4}{c}{\huge NDCG} 
& \multicolumn{4}{c}{\huge Recall} & \multicolumn{4}{c}{\huge NDCG} 
& \multicolumn{4}{c}{\huge Recall} & \multicolumn{4}{c}{\huge NDCG}\\
& \huge@5 & \huge@10 & \huge@20 & \huge@50 & \huge@5 & \huge@10 & \huge@20 & \huge@50
& \huge@5 & \huge@10 & \huge@20 & \huge@50 & \huge@5 & \huge@10 & \huge@20 & \huge@50
& \huge@5 & \huge@10 & \huge@20 & \huge@50 & \huge@5 & \huge@10 & \huge@20 & \huge@50 \\
\midrule
\huge F-SDRM       
& \makecell{\LARGE ${0.1211}$ \\ $\pm 0.0100$} & \makecell{\LARGE ${0.1697}$ \\ $\pm 0.0111$} & \makecell{\LARGE ${0.2440}$ \\ $\pm 0.0180$} & \makecell{\LARGE ${0.3399}$ \\ $\pm 0.0279$} & \makecell{\LARGE ${0.0830}$ \\ $\pm 0.0064$} & \makecell{\LARGE ${0.0986}$ \\ $\pm 0.0065$} & \makecell{\LARGE ${0.1175}$ \\ $\pm 0.0083$} & \makecell{\LARGE ${0.1367}$ \\ $\pm 0.0101$}& \makecell{\LARGE ${0.3043}$ \\ $\pm 0.0027$} & \makecell{\LARGE ${0.3210}$ \\ $\pm 0.0123$} & \makecell{\LARGE ${0.3403}$ \\ $\pm 0.0162$} & \makecell{\LARGE ${0.3909}$ \\ $\pm 0.0143$}& \makecell{\LARGE ${0.2981}$ \\ $\pm 0.0109$} & \makecell{\LARGE ${0.3040}$ \\ $\pm 0.0081$} & \makecell{\LARGE ${0.3102}$ \\ $\pm 0.0051$} & \makecell{\LARGE ${0.3243}$ \\ $\pm 0.0050$}& \makecell{\LARGE ${0.2988}$ \\ $\pm 0.0050$} & \makecell{\LARGE ${0.2846}$ \\ $\pm 0.0044$} & \makecell{\LARGE ${0.3025}$ \\ $\pm 0.0041$} & \makecell{\LARGE ${0.3926}$ \\ $\pm 0.0040$}& \makecell{\LARGE ${0.3228}$ \\ $\pm 0.0042$} & \makecell{\LARGE ${0.3038}$ \\ $\pm 0.0069$} & \makecell{\LARGE ${0.2989}$ \\ $\pm 0.0028$} & \makecell{\LARGE ${0.3290}$ \\ $\pm 0.0012$}\\
\bottomrule
\end{tabular}
}
\end{table*}

\begin{table*}[htbp]
\centering
\caption{Performance comparison against generative baselines on \textbf{NeuMF} recommendation algorithm.}
\resizebox{\textwidth}{!}{
\renewcommand{\arraystretch}{2.5}
\begin{tabular}{lcccccccccccccccccccccccc}
\toprule
\multicolumn{8}{c}{\hspace{15em}\huge ProgrammableWeb} &
\multicolumn{8}{c}{\hspace{15em}\huge Amazon Luxury Beauty} & 
\multicolumn{8}{c}{\hspace{15em}\huge MovieLens 100K} \\
\cmidrule(r){2-9} \cmidrule(r){10-17} \cmidrule(r){18-25}
\textbf{\huge Method} & \multicolumn{4}{c}{\huge Recall} & \multicolumn{4}{c}{\huge NDCG} 
& \multicolumn{4}{c}{\huge Recall} & \multicolumn{4}{c}{\huge NDCG} 
& \multicolumn{4}{c}{\huge Recall} & \multicolumn{4}{c}{\huge NDCG}\\
& \huge@5 & \huge@10 & \huge@20 & \huge@50 & \huge@5 & \huge@10 & \huge@20 & \huge@50
& \huge@5 & \huge@10 & \huge@20 & \huge@50 & \huge@5 & \huge@10 & \huge@20 & \huge@50
& \huge@5 & \huge@10 & \huge@20 & \huge@50 & \huge@5 & \huge@10 & \huge@20 & \huge@50 \\
\midrule
\huge F-SDRM       
& \makecell{\LARGE ${0.2571}$ \\ $\pm 0.0147$} & \makecell{\LARGE ${0.3607}$ \\ $\pm 0.0450$} & \makecell{\LARGE ${0.4447}$ \\ $\pm 0.0040$} & \makecell{\LARGE ${0.6208}$ \\ $\pm 0.0128$} & \makecell{\LARGE ${0.0397}$ \\ $\pm 0.0015$} & \makecell{\LARGE ${0.0312}$ \\ $\pm 0.0027$} & \makecell{\LARGE ${0.0223}$ \\ $\pm 0.0008$} & \makecell{\LARGE ${0.0141}$ \\ $\pm 0.0002$}& \makecell{\LARGE ${0.3164}$ \\ $\pm 0.0019$} & \makecell{\LARGE ${0.3486}$ \\ $\pm 0.0039$} & \makecell{\LARGE ${0.3821}$ \\ $\pm 0.0066$} & \makecell{\LARGE ${0.4350}$ \\ $\pm 0.0087$}& \makecell{\LARGE ${0.1682}$ \\ $\pm 0.0050$} & \makecell{\LARGE ${0.1153}$ \\ $\pm 0.0028$} & \makecell{\LARGE ${0.0772}$ \\ $\pm 0.0018$} & \makecell{\LARGE ${0.0443}$ \\ $\pm 0.0011$}& \makecell{\LARGE ${0.1154}$ \\ $\pm 0.0214$} & \makecell{\LARGE ${0.1237}$ \\ $\pm 0.0174$} & \makecell{\LARGE ${0.1539}$ \\ $\pm 0.0124$} & \makecell{\LARGE ${0.2732}$ \\ $\pm 0.0308$}& \makecell{\LARGE ${0.1119}$ \\ $\pm 0.0191$} & \makecell{\LARGE ${0.1002}$ \\ $\pm 0.0194$} & \makecell{\LARGE ${0.0824}$ \\ $\pm 0.0148$} & \makecell{\LARGE ${0.0653}$ \\ $\pm 0.0114$}\\
\bottomrule
\end{tabular}
}
\end{table*}

\begin{table*}[htbp]
\centering
\caption{Performance comparison against generative baselines on \textbf{NGCF} recommendation algorithm.}
\resizebox{\textwidth}{!}{
\renewcommand{\arraystretch}{2.5}
\begin{tabular}{lcccccccccccccccccccccccc}
\toprule
\multicolumn{8}{c}{\hspace{15em}\huge ProgrammableWeb} &
\multicolumn{8}{c}{\hspace{15em}\huge Amazon Luxury Beauty} & 
\multicolumn{8}{c}{\hspace{15em}\huge MovieLens 100K} \\
\cmidrule(r){2-9} \cmidrule(r){10-17} \cmidrule(r){18-25}
\textbf{\huge Method} & \multicolumn{4}{c}{\huge Recall} & \multicolumn{4}{c}{\huge NDCG} 
& \multicolumn{4}{c}{\huge Recall} & \multicolumn{4}{c}{\huge NDCG} 
& \multicolumn{4}{c}{\huge Recall} & \multicolumn{4}{c}{\huge NDCG}\\
& \huge@5 & \huge@10 & \huge@20 & \huge@50 & \huge@5 & \huge@10 & \huge@20 & \huge@50
& \huge@5 & \huge@10 & \huge@20 & \huge@50 & \huge@5 & \huge@10 & \huge@20 & \huge@50
& \huge@5 & \huge@10 & \huge@20 & \huge@50 & \huge@5 & \huge@10 & \huge@20 & \huge@50 \\
\midrule
\huge F-SDRM       
& \makecell{\LARGE ${0.2520}$ \\ $\pm 0.0320$} & \makecell{\LARGE ${0.3292}$ \\ $\pm 0.0115$} & \makecell{\LARGE ${0.3875}$ \\ $\pm 0.0437$} & \makecell{\LARGE ${0.4815}$ \\ $\pm 0.0769$} & \makecell{\LARGE ${0.0411}$ \\ $\pm 0.0039$} & \makecell{\LARGE ${0.0313}$ \\ $\pm 0.0001$} & \makecell{\LARGE ${0.0218}$ \\ $\pm 0.0010$} & \makecell{\LARGE ${0.0131}$ \\ $\pm 0.0010$}& \makecell{\LARGE ${0.3210}$ \\ $\pm 0.0180$} & \makecell{\LARGE ${0.3535}$ \\ $\pm 0.0152$} & \makecell{\LARGE ${0.3836}$ \\ $\pm 0.0048$} & \makecell{\LARGE ${0.4455}$ \\ $\pm 0.0123$}& \makecell{\LARGE ${0.1632}$ \\ $\pm 0.0088$} & \makecell{\LARGE ${0.1095}$ \\ $\pm 0.0057$} & \makecell{\LARGE ${0.0727}$ \\ $\pm 0.0032$} & \makecell{\LARGE ${0.0415}$ \\ $\pm 0.0017$}& \makecell{\LARGE ${0.2189}$ \\ $\pm 0.0264$} & \makecell{\LARGE ${0.2287}$ \\ $\pm 0.0162$} & \makecell{\LARGE ${0.2820}$ \\ $\pm 0.0084$} & \makecell{\LARGE ${0.4296}$ \\ $\pm 0.0262$}& \makecell{\LARGE ${0.2207}$ \\ $\pm 0.0260$} & \makecell{\LARGE ${0.1902}$ \\ $\pm 0.0167$} & \makecell{\LARGE ${0.1584}$ \\ $\pm 0.0112$} & \makecell{\LARGE ${0.1151}$ \\ $\pm 0.0074$}\\
\bottomrule
\end{tabular}
}
\end{table*}

\begin{table*}[htbp]
\centering
\caption{Performance comparison against generative baselines on \textbf{PureSVD} recommendation algorithm.}
\resizebox{\textwidth}{!}{
\renewcommand{\arraystretch}{2.5}
\begin{tabular}{lcccccccccccccccccccccccc}
\toprule
\multicolumn{8}{c}{\hspace{15em}\huge ProgrammableWeb} &
\multicolumn{8}{c}{\hspace{15em}\huge Amazon Luxury Beauty} & 
\multicolumn{8}{c}{\hspace{15em}\huge MovieLens 100K} \\
\cmidrule(r){2-9} \cmidrule(r){10-17} \cmidrule(r){18-25}
\textbf{\huge Method} & \multicolumn{4}{c}{\huge Recall} & \multicolumn{4}{c}{\huge NDCG} 
& \multicolumn{4}{c}{\huge Recall} & \multicolumn{4}{c}{\huge NDCG} 
& \multicolumn{4}{c}{\huge Recall} & \multicolumn{4}{c}{\huge NDCG}\\
& \huge@5 & \huge@10 & \huge@20 & \huge@50 & \huge@5 & \huge@10 & \huge@20 & \huge@50
& \huge@5 & \huge@10 & \huge@20 & \huge@50 & \huge@5 & \huge@10 & \huge@20 & \huge@50
& \huge@5 & \huge@10 & \huge@20 & \huge@50 & \huge@5 & \huge@10 & \huge@20 & \huge@50 \\
\midrule
\huge F-SDRM       
& \makecell{\LARGE ${0.0839}$ \\ $\pm 0.0034$} & \makecell{\LARGE ${0.1463}$ \\ $\pm 0.0090$} & \makecell{\LARGE ${0.2182}$ \\ $\pm 0.0045$} & \makecell{\LARGE ${0.3195}$ \\ $\pm 0.0017$} & \makecell{\LARGE ${0.0560}$ \\ $\pm 0.0021$} & \makecell{\LARGE ${0.0757}$ \\ $\pm 0.0036$} & \makecell{\LARGE ${0.0943}$ \\ $\pm 0.0022$} & \makecell{\LARGE ${0.1142}$ \\ $\pm 0.0013$}& \makecell{\LARGE ${0.2887}$ \\ $\pm 0.0060$} & \makecell{\LARGE ${0.3215}$ \\ $\pm 0.0057$} & \makecell{\LARGE ${0.3389}$ \\ $\pm 0.0025$} & \makecell{\LARGE ${0.3863}$ \\ $\pm 0.0092$}& \makecell{\LARGE ${0.2800}$ \\ $\pm 0.0019$} & \makecell{\LARGE ${0.2931}$ \\ $\pm 0.0015$} & \makecell{\LARGE ${0.2992}$ \\ $\pm 0.0004$} & \makecell{\LARGE ${0.3122}$ \\ $\pm 0.0020$}& \makecell{\LARGE ${0.3196}$ \\ $\pm 0.0080$} & \makecell{\LARGE ${0.2908}$ \\ $\pm 0.0066$} & \makecell{\LARGE ${0.3017}$ \\ $\pm 0.0106$} & \makecell{\LARGE ${0.4061}$ \\ $\pm 0.0090$}& \makecell{\LARGE ${0.3367}$ \\ $\pm 0.0085$} & \makecell{\LARGE ${0.3095}$ \\ $\pm 0.0081$} & \makecell{\LARGE ${0.3023}$ \\ $\pm 0.0096$} & \makecell{\LARGE ${0.3388}$ \\ $\pm 0.0086$}\\
\bottomrule
\end{tabular}
}
\end{table*}

\begin{table*}[htbp]
\centering
\caption{Performance comparison against generative baselines on \textbf{NGSGL} recommendation algorithm.}
\resizebox{\textwidth}{!}{
\renewcommand{\arraystretch}{2.5}
\begin{tabular}{lcccccccccccccccccccccccc}
\toprule
\multicolumn{8}{c}{\hspace{15em}\huge ProgrammableWeb} &
\multicolumn{8}{c}{\hspace{15em}\huge Amazon Luxury Beauty} & 
\multicolumn{8}{c}{\hspace{15em}\huge MovieLens 100K} \\
\cmidrule(r){2-9} \cmidrule(r){10-17} \cmidrule(r){18-25}
\textbf{\huge Method} & \multicolumn{4}{c}{\huge Recall} & \multicolumn{4}{c}{\huge NDCG} 
& \multicolumn{4}{c}{\huge Recall} & \multicolumn{4}{c}{\huge NDCG} 
& \multicolumn{4}{c}{\huge Recall} & \multicolumn{4}{c}{\huge NDCG}\\
& \huge@5 & \huge@10 & \huge@20 & \huge@50 & \huge@5 & \huge@10 & \huge@20 & \huge@50
& \huge@5 & \huge@10 & \huge@20 & \huge@50 & \huge@5 & \huge@10 & \huge@20 & \huge@50
& \huge@5 & \huge@10 & \huge@20 & \huge@50 & \huge@5 & \huge@10 & \huge@20 & \huge@50 \\
\midrule
\huge F-SDRM       
& \makecell{\LARGE ${0.2246}$ \\ $\pm 0.0295$} & \makecell{\LARGE ${0.3287}$ \\ $\pm 0.0316$} & \makecell{\LARGE ${0.4166}$ \\ $\pm 0.0167$} & \makecell{\LARGE ${0.5378}$ \\ $\pm 0.0572$} & \makecell{\LARGE ${0.0389}$ \\ $\pm 0.0030$} & \makecell{\LARGE ${0.0313}$ \\ $\pm 0.0009$} & \makecell{\LARGE ${0.0228}$ \\ $\pm 0.0005$} & \makecell{\LARGE ${0.0140}$ \\ $\pm 0.0009$}& \makecell{\LARGE ${0.3257}$ \\ $\pm 0.0088$} & \makecell{\LARGE ${0.3370}$ \\ $\pm 0.0126$} & \makecell{\LARGE ${0.3569}$ \\ $\pm 0.0135$} & \makecell{\LARGE ${0.4175}$ \\ $\pm 0.0176$}& \makecell{\LARGE ${0.1628}$ \\ $\pm 0.0051$} & \makecell{\LARGE ${0.1078}$ \\ $\pm 0.0041$} & \makecell{\LARGE ${0.0709}$ \\ $\pm 0.0024$} & \makecell{\LARGE ${0.0404}$ \\ $\pm 0.0009$}& \makecell{\LARGE ${0.2083}$ \\ $\pm 0.0158$} & \makecell{\LARGE ${0.2207}$ \\ $\pm 0.0030$} & \makecell{\LARGE ${0.2535}$ \\ $\pm 0.0066$} & \makecell{\LARGE ${0.3848}$ \\ $\pm 0.0094$}& \makecell{\LARGE ${0.2136}$ \\ $\pm 0.0130$} & \makecell{\LARGE ${0.1864}$ \\ $\pm 0.0044$} & \makecell{\LARGE ${0.1496}$ \\ $\pm 0.0050$} & \makecell{\LARGE ${0.1092}$ \\ $\pm 0.0045$}\\
\bottomrule
\end{tabular}
}
\end{table*}

\begin{table*}[htbp]
\centering
\caption{Performance comparison against generative baselines on \textbf{NGsimGCL} recommendation algorithm.}
\resizebox{\textwidth}{!}{
\renewcommand{\arraystretch}{2.5}
\begin{tabular}{lcccccccccccccccccccccccc}
\toprule
\multicolumn{8}{c}{\hspace{15em}\huge ProgrammableWeb} &
\multicolumn{8}{c}{\hspace{15em}\huge Amazon Luxury Beauty} & 
\multicolumn{8}{c}{\hspace{15em}\huge MovieLens 100K} \\
\cmidrule(r){2-9} \cmidrule(r){10-17} \cmidrule(r){18-25}
\textbf{\huge Method} & \multicolumn{4}{c}{\huge Recall} & \multicolumn{4}{c}{\huge NDCG} 
& \multicolumn{4}{c}{\huge Recall} & \multicolumn{4}{c}{\huge NDCG} 
& \multicolumn{4}{c}{\huge Recall} & \multicolumn{4}{c}{\huge NDCG}\\
& \huge@5 & \huge@10 & \huge@20 & \huge@50 & \huge@5 & \huge@10 & \huge@20 & \huge@50
& \huge@5 & \huge@10 & \huge@20 & \huge@50 & \huge@5 & \huge@10 & \huge@20 & \huge@50
& \huge@5 & \huge@10 & \huge@20 & \huge@50 & \huge@5 & \huge@10 & \huge@20 & \huge@50 \\
\midrule
\huge F-SDRM       
& \makecell{\LARGE ${0.2422}$ \\ $\pm 0.0260$} & \makecell{\LARGE ${0.3017}$ \\ $\pm 0.0267$} & \makecell{\LARGE ${0.3510}$ \\ $\pm 0.0291$} & \makecell{\LARGE ${0.4603}$ \\ $\pm 0.0231$} & \makecell{\LARGE ${0.0425}$ \\ $\pm 0.0011$} & \makecell{\LARGE ${0.0315}$ \\ $\pm 0.0005$} & \makecell{\LARGE ${0.0219}$ \\ $\pm 0.0004$} & \makecell{\LARGE ${0.0134}$ \\ $\pm 0.0006$}& \makecell{\LARGE ${0.3234}$ \\ $\pm 0.0146$} & \makecell{\LARGE ${0.3556}$ \\ $\pm 0.0123$} & \makecell{\LARGE ${0.3722}$ \\ $\pm 0.0147$} & \makecell{\LARGE ${0.4624}$ \\ $\pm 0.0155$}& \makecell{\LARGE ${0.1613}$ \\ $\pm 0.0046$} & \makecell{\LARGE ${0.1089}$ \\ $\pm 0.0026$} & \makecell{\LARGE ${0.0713}$ \\ $\pm 0.0016$} & \makecell{\LARGE ${0.0411}$ \\ $\pm 0.0010$}& \makecell{\LARGE ${0.2011}$ \\ $\pm 0.0258$} & \makecell{\LARGE ${0.2154}$ \\ $\pm 0.0193$} & \makecell{\LARGE ${0.2615}$ \\ $\pm 0.0242$} & \makecell{\LARGE ${0.3823}$ \\ $\pm 0.0230$}& \makecell{\LARGE ${0.2006}$ \\ $\pm 0.0205$} & \makecell{\LARGE ${0.1730}$ \\ $\pm 0.0180$} & \makecell{\LARGE ${0.1421}$ \\ $\pm 0.0130$} & \makecell{\LARGE ${0.1026}$ \\ $\pm 0.0080$}\\
\bottomrule
\end{tabular}
}
\end{table*}

\begin{table*}[htbp]
\centering
\caption{Performance comparison against generative baselines on \textbf{HybridRec} recommendation algorithm.}
\resizebox{\textwidth}{!}{
\renewcommand{\arraystretch}{2.5}
\begin{tabular}{lcccccccccccccccccccccccc}
\toprule
\multicolumn{8}{c}{\hspace{15em}\huge ProgrammableWeb} &
\multicolumn{8}{c}{\hspace{15em}\huge Amazon Luxury Beauty} & 
\multicolumn{8}{c}{\hspace{15em}\huge MovieLens 100K} \\
\cmidrule(r){2-9} \cmidrule(r){10-17} \cmidrule(r){18-25}
\textbf{\huge Method} & \multicolumn{4}{c}{\huge Recall} & \multicolumn{4}{c}{\huge NDCG} 
& \multicolumn{4}{c}{\huge Recall} & \multicolumn{4}{c}{\huge NDCG} 
& \multicolumn{4}{c}{\huge Recall} & \multicolumn{4}{c}{\huge NDCG}\\
& \huge@5 & \huge@10 & \huge@20 & \huge@50 & \huge@5 & \huge@10 & \huge@20 & \huge@50
& \huge@5 & \huge@10 & \huge@20 & \huge@50 & \huge@5 & \huge@10 & \huge@20 & \huge@50
& \huge@5 & \huge@10 & \huge@20 & \huge@50 & \huge@5 & \huge@10 & \huge@20 & \huge@50 \\
\midrule
\huge F-SDRM       
& \makecell{\LARGE ${0.1061}$ \\ $\pm 0.0115$} & \makecell{\LARGE ${0.1709}$ \\ $\pm 0.0076$} & \makecell{\LARGE ${0.2386}$ \\ $\pm 0.0086$} & \makecell{\LARGE ${0.3387}$ \\ $\pm 0.0181$} & \makecell{\LARGE ${0.0676}$ \\ $\pm 0.0050$} & \makecell{\LARGE ${0.0883}$ \\ $\pm 0.0047$} & \makecell{\LARGE ${0.1055}$ \\ $\pm 0.0045$} & \makecell{\LARGE ${0.1254}$ \\ $\pm 0.0066$}& \makecell{\LARGE ${0.3024}$ \\ $\pm 0.0049$} & \makecell{\LARGE ${0.3292}$ \\ $\pm 0.0046$} & \makecell{\LARGE ${0.3534}$ \\ $\pm 0.0123$} & \makecell{\LARGE ${0.3924}$ \\ $\pm 0.0189$}& \makecell{\LARGE ${0.2945}$ \\ $\pm 0.0045$} & \makecell{\LARGE ${0.3046}$ \\ $\pm 0.0013$} & \makecell{\LARGE ${0.3128}$ \\ $\pm 0.0009$} & \makecell{\LARGE ${0.3236}$ \\ $\pm 0.0037$}& \makecell{\LARGE ${0.3300}$ \\ $\pm 0.0091$} & \makecell{\LARGE ${0.2972}$ \\ $\pm 0.0154$} & \makecell{\LARGE ${0.3039}$ \\ $\pm 0.0038$} & \makecell{\LARGE ${0.4137}$ \\ $\pm 0.0046$}& \makecell{\LARGE ${0.3513}$ \\ $\pm 0.0127$} & \makecell{\LARGE ${0.3199}$ \\ $\pm 0.0165$} & \makecell{\LARGE ${0.3075}$ \\ $\pm 0.0071$} & \makecell{\LARGE ${0.3444}$ \\ $\pm 0.0070$}\\
\bottomrule
\end{tabular}
}
\label{tab:HybridRec}
\end{table*}

\section{Full Ablation Study Results}\label{Full Ablation}
\begin{figure}[h]
\centering
\includegraphics[width=\linewidth,]{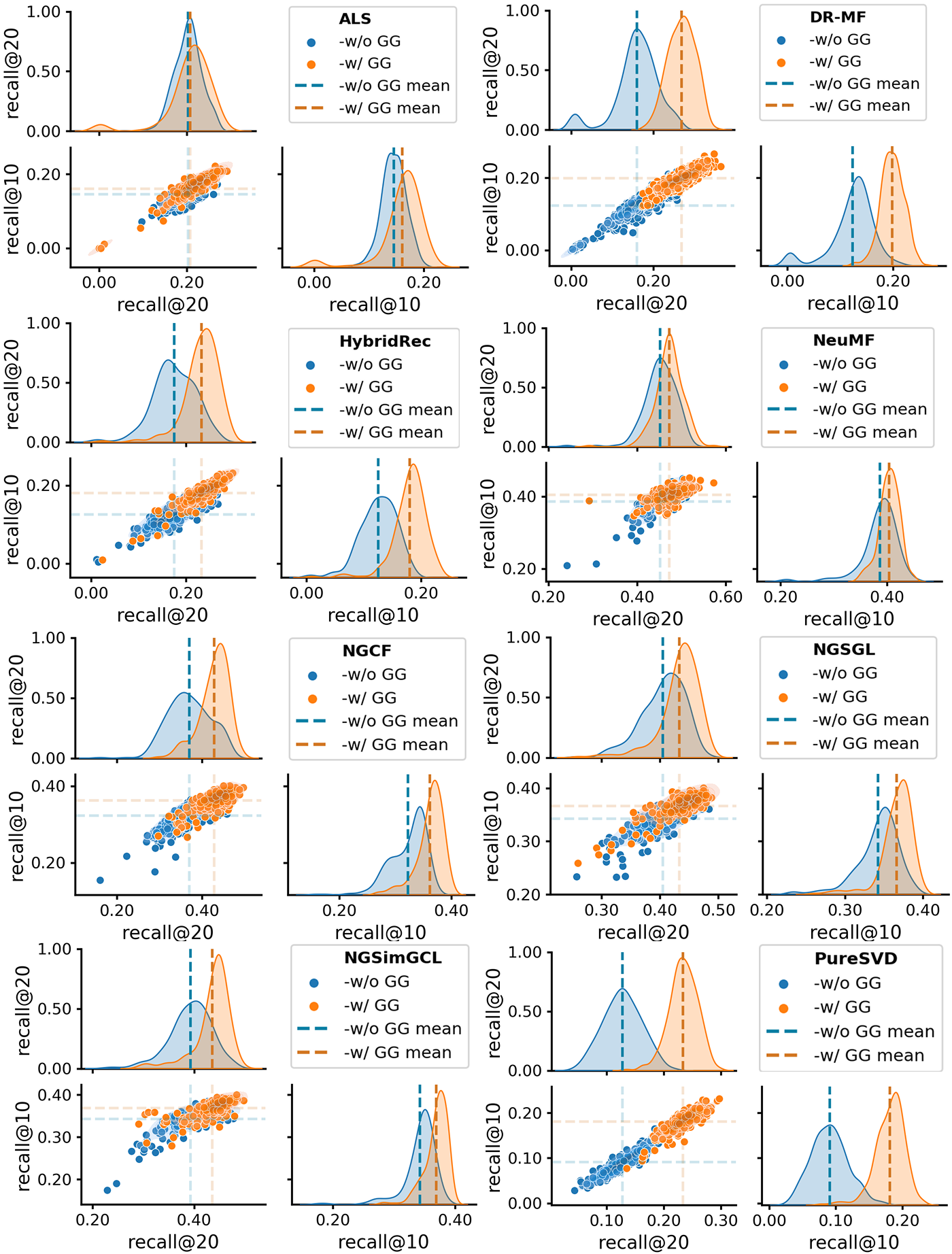}
\caption{Paired Recall@10 and Recall@20 distributions on the ProgrammableWeb dataset.}
\label{distribution plot all}
\end{figure}

To assess the contribution of individual components, we conduct a comprehensive ablation study on the two core modules: the node-level graph generation component, GG, and the structural denoising preference modeling component, SD. Table~\ref{tab:full ablation} presents the full ablation results, and Figure~\ref{distribution plot all} shows the full paired Recall@10 and Recall@20 distributions on the ProgrammableWeb dataset, with and without the node-level graph generation component.

\begin{table*}[htbp]
\centering
\caption{Full ablation performance table.}
\label{tab:ablation_comparison}
\resizebox{\textwidth}{!}{
\renewcommand{\arraystretch}{3.0}
\begin{tabular}{l|cccc|cccc|cccc|cccc|cccc|cccc}
\toprule
& \multicolumn{8}{c|}{\huge\textbf{ProgrammableWeb}} 
& \multicolumn{8}{c|}{\huge\textbf{Amazon Luxury Beauty}} 
& \multicolumn{8}{c}{\huge\textbf{MovieLens 100K}} \\
\cmidrule(r){2-9} \cmidrule(r){10-17} \cmidrule(r){18-25}
\huge{Algorithm} 
& \multicolumn{4}{c|}{\huge Recall} & \multicolumn{4}{c|}{\huge NDCG}
& \multicolumn{4}{c|}{\huge Recall} & \multicolumn{4}{c|}{\huge NDCG}
& \multicolumn{4}{c|}{\huge Recall} & \multicolumn{4}{c}{\huge NDCG} \\
\cmidrule(r){2-5} \cmidrule(r){6-9}
\cmidrule(r){10-13} \cmidrule(r){14-17}
\cmidrule(r){18-21} \cmidrule(r){22-25}
& {\huge @5}  & {\huge @10}  & {\huge @20}  & {\huge @50}
& {\huge @5}  & {\huge @10}  & {\huge @20}  & {\huge @50}
& {\huge @5}  & {\huge @10}  & {\huge @20}  & {\huge @50}
& {\huge @5}  & {\huge @10}  & {\huge @20}  & {\huge @50}
& {\huge @5}  & {\huge @10}  & {\huge @20}  & {\huge @50}
& {\huge @5}  & {\huge @10}  & {\huge @20}  & {\huge @50} \\
\bottomrule
\end{tabular}
}
\label{tab:full ablation}
\end{table*}

\section{Case Study Details}\label{Case study improvement}
To investigate how NodeDiffRec benefits recommendation, we visualize user-item interactions around two representative users before and after augmentation. Table~\ref{metrics for case study} reports exact Recall@k and NDCG@k of the selected users: User\_861 and User\_932, where significant improvement can be observed.

\begin{table}[h]
\centering
\renewcommand{\arraystretch}{0.3}
\caption{Recommendation metrics for two users: User\_861 and User\_932 before and after the augmentation from the proposed NodeDiffRec.}
\resizebox{\linewidth}{!}{
\begin{tabular}{llcccc}
\toprule
\textbf{User} & \textbf{Metric} & \textbf{@5} & \textbf{@10} & \textbf{@20} & \textbf{@50} \\
\midrule
\multirow{4}{*}{User\_861} & Recall-bef. & 0.200 & 0.300 & 0.200 & 0.333 \\
                           & Recall-aft. & \textbf{0.600} & \textbf{0.600} & \textbf{0.350} & \textbf{0.429} \\
\cmidrule(lr){2-6}
                           & NDCG-bef.   & 0.214 & 0.268 & 0.206 & 0.160 \\
                           & NDCG-aft.   & \textbf{0.684} & \textbf{0.653} & \textbf{0.456} & \textbf{0.281} \\
\midrule
\multirow{4}{*}{User\_932} & Recall-bef. & 0.000 & 0.500 & 0.500 & 0.500 \\
                           & Recall-aft. & \textbf{0.500} & \textbf{0.500} & \textbf{1.000} & \textbf{1.000} \\
\cmidrule(lr){2-6}
                           & NDCG-bef.   & 0.000 & 0.078 & 0.050 & 0.027 \\
                           & NDCG-aft.   & \textbf{0.214} & \textbf{0.139} & \textbf{0.128} & \textbf{0.070} \\
\bottomrule
\end{tabular}
}
\label{metrics for case study}
\end{table}

\end{document}